\documentclass[11pt]{article}
\usepackage{amsmath,amsfonts,amssymb,bm}

\usepackage{graphicx}

\makeatletter
\@addtoreset{equation}{section}
\makeatother

\makeatletter
\long\def\@makecaption#1#2{{\small
\advance\leftskip1cm
\advance\rightskip1cm
\vskip\abovecaptionskip
\sbox\@tempboxa{#1: #2}%
\ifdim \wd\@tempboxa >\hsize
 #1: #2\par
\else
\global \@minipagefalse
\hb@xt@\hsize{\hfil\box\@tempboxa\hfil}%
\fi
\vskip\belowcaptionskip}}
\makeatother

\begin{document}

\noindent
{\bf\Large 
Geometrical Excess Entropy Production in Nonequilibrium Quantum Systems
}
\bigskip

\noindent
{\bf 
Tatsuro Yuge%
\footnote{Department of Physics, Osaka University, Machikaneyama-Cho, Toyonaka, Osaka, 560-0043, Japan}, 
Takahiro Sagawa%
\footnote{The Hakubi Center, Kyoto University, Yosida Ushinomiya-Cho, Sakyo-ku, Kyoto, 606-8302, Japan}$^,$%
\footnote{Yukawa Institute for Theoretical Physics, Kyoto University, Kitashirakawa Oiwake-Cho, Sakyo-ku, Kyoto, 606-8502, Japan}$^,$%
\footnote{Present address: Department of Basic Science, The University of Tokyo, Komaba, Meguro-ku, Tokyo, 153-8902, Japan},
Ayumu Sugita%
\footnote{Department of Applied Physics, Osaka City University, Sugimoto, Osaka, 558-8585, Japan},
and 
Hisao Hayakawa$^3$
}
\begin{center}
\today
\end{center}

\begin{abstract}
For open systems described by the quantum Markovian master equation, 
we study a possible extension of the Clausius equality to 
quasistatic operations between nonequilibrium steady states (NESSs). 
We investigate the excess heat divided by temperature (i.e., excess entropy production) 
which is transferred into the system during the operations. 
We derive a geometrical expression for the excess entropy production, 
which is analogous to the Berry phase in unitary evolution. 
Our result implies that in general one cannot define a scalar potential 
whose difference coincides with the excess entropy production in a thermodynamic process, 
and that a vector potential plays a crucial role in the thermodynamics for NESSs. 
In the weakly nonequilibrium regime, we show that the geometrical expression reduces to 
the extended Clausius equality derived by Saito and Tasaki (J. Stat. Phys. {\bf 145}, 1275 (2011)). 
As an example, we investigate a spinless electron system in quantum dots. 
We find that one can define a scalar potential when the parameters of only one of the reservoirs are modified 
in a non-interacting system, but this is no longer the case for an interacting system.
\end{abstract}

\section{Introduction}

\subsection{Background and Motivation}

Thermodynamics and statistical mechanics are universal and powerful frameworks 
to describe systems in equilibrium states. 
In equilibrium thermodynamics, the central quantity is the entropy $S$, 
which describes both the macroscopic properties of equilibrium systems 
and the fundamental limits on the possible transitions among the equilibrium states. 
Its operational definition relies on the Clausius equality: 
\begin{align}
\Delta S = \beta Q. 
\label{equilibriumClausius}
\end{align}
This equality is valid for quasistatic operations between two equilibrium states. 
Here, $\Delta S$ is the change in the entropy of the system before and after the operation, 
$\beta$ is the inverse temperature of the reservoir that is in contact with the system, 
and $Q$ is the heat transferred from the reservoir to the system during the operation.
Equilibrium statistical mechanics tells that the entropy $S$ is given by the Shannon entropy of 
the probability distribution (von Neumann entropy of the density matrix) of microscopic states 
in the equilibrium classical (quantum) system.
This connects the microscopic physics to the macroscopic one, 
where we do not need to solve the equation of motion in the microscopic level.

The construction of analogous frameworks of thermodynamics and statistical mechanics for nonequilibrium systems 
has been one of the central subjects in statistical physics 
\cite{Bertini_etal_2,GrootMazur,Landauer,OonoPaniconi,Prigogine,SasaTasaki,SpohnLebowitz,Zubarev}.
Recently there has been progress in the extension of the Clausius equality 
to nonequilibrium steady states (NESSs) \cite{KNST1,KNST2,Ruelle,SaitoTasaki} 
(see also Refs.~\cite{Bertini_etal_3,Bertini_etal_4,DeffnerLutz,HatanoSasa,Nakagawa,Takara}).
In these studies, the excess heat $Q^{\rm ex}$, proposed in Ref. \cite{OonoPaniconi}, 
has been used instead of the total heat $Q$ in the equilibrium equality (\ref{equilibriumClausius}). 
The excess heat $Q^{\rm ex}$ is defined by subtracting from $Q$ the contribution $Q_{\rm hk}$ 
(called housekeeping heat) of steady heat dissipation in NESS.
Then it has been shown that in the weakly nonequilibrium regime there exists a scalar potential $S_{\rm sym}$
which satisfies the extended Clausius equality, 
\begin{align}
\Delta S_{\rm sym} = \beta Q^{\rm ex}, 
\label{extendedClausius}
\end{align}
for quasistatic operations.
We refer to $\beta Q^{\rm ex}$ as excess entropy production during the operation.
Moreover, $S_{\rm sym}$ in Eq.~(\ref{extendedClausius}) is given by a symmetrized version of 
the Shannon (von Neumann) entropy of the NESS.

More recently, a formula for the excess entropy production during quasistatic operations 
in the strongly nonequilibrium regime has been derived in generic classical systems 
described by the Markov jump process \cite{SagawaHayakawa}.
This formula is expressed by a geometrical (Berry-phase-like) quantity \cite{Berry,SinitsynNemenman2007a}; 
i.e., the excess entropy production is given by a line integral of a vector potential 
in the operation parameter space.
This implies that, in general, the extended Clausius equality does not hold in the strongly nonequilibrium regime. 
Therefore the operational definition of the nonequilibrium entropy (scalar thermodynamic potential) 
through the excess entropy production is impossible, 
but the vector potential plays an important role in the thermodynamics for NESSs.

Since the extended Clausius equality holds for both the classical systems \cite{KNST1,KNST2} 
and quantum systems \cite{SaitoTasaki}, 
it is expected that the geometrical expression for the excess entropy production 
in the strongly nonequilibrium regime \cite{SagawaHayakawa} can also be generalized to quantum systems.
In this paper we show that this expectation is true; 
we derive a quantum version of the geometrical expression in open systems 
described by the quantum Markovian master equation (QMME).
This result suggests the universality of the geometrical expression 
and the importance of the vector potential in the thermodynamics for NESSs.
It should be noted that in the field of adiabatic pumping 
the path-dependent physical variables play important roles thanks to the existence of the geometrical phase, 
where there exists a current between reservoirs without dc bias 
\cite{Brouwer,Kouwenhoven_etal,SinitsynNemenman2007a,Switkes_etal,Thouless,Yuge_etal}. 
We believe that the concept of path-dependent quantities is important 
in general nonequilibrium situations.

\begin{figure}[bt]
\begin{center}
\includegraphics[width=0.95\linewidth]{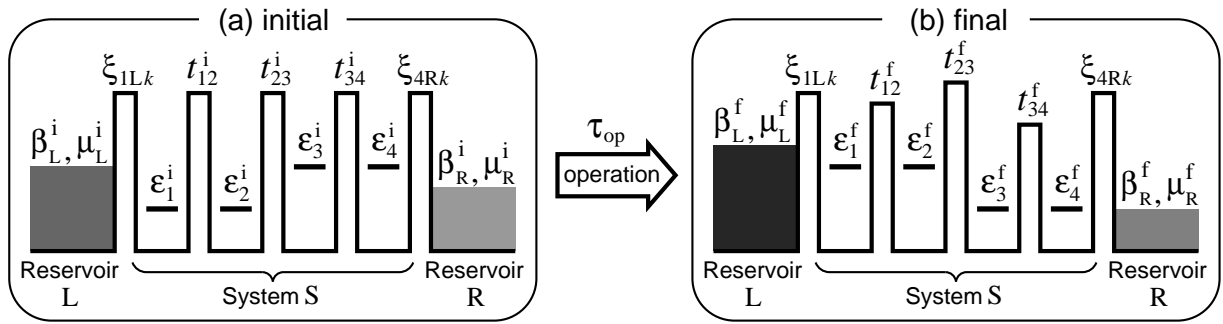}
\caption{
Example of a setting of nonequilibrium system and thermodynamic operation onto it. 
A series of quantum dots (system S) are connected to electron reservoirs. 
The energy levels of the dots are denoted by $\varepsilon_i$ and 
the transition probability amplitudes between dots by $t_{ii'}$. 
The reservoirs are in equilibrium states with inverse temperatures $\{\beta_b\}_b$ 
and chemical potentials $\{\mu_b\}_b$. 
By changing these parameters from the initial values (a) to the final ones (b) in a time scale $\tau_{\rm op}$, 
one can perform a thermodynamic operation on the system S.
}
\label{fig:example}
\end{center}
\end{figure}

\subsection{Brief Summary of Geometrical Expression for Excess Entropy Production}\label{sec:briefSummary}

Here we briefly explain our setting of a nonequilibrium system,  
the quasistatic operation between NESSs, and the excess entropy production. 
We also present a brief summary of the main result, 
the geometrical expression for the excess entropy production. 

Suppose that a quantum system S is weakly coupled to multiple reservoirs 
with definite inverse temperatures $\{\beta_b\}_b$ and chemical potentials $\{\mu_b\}_b$. 
An example of the setup is given in Fig.~\ref{fig:example}, 
where S is composed of quantum dots and connected to two electron reservoirs 
(see Sec.~\ref{sec:examples} for the results in this example). 
The system S is characterized by a set of parameters $\bm{\alpha}$, 
which includes system internal parameters $\bm{\alpha}_{\rm S}$ 
(e.g., dot levels and tunnel barriers in Fig.~\ref{fig:example}) 
and reservoir parameters $\bm{\alpha}_{\rm B}=\{\beta_b,\mu_b\}_b$. 
One can perform a thermodynamic operation by changing the parameters $\bm{\alpha}$ in time. 
The reservoirs are assumed to remain in the equilibrium states during the operation.

In this setting, the dynamics of the system S can be described by the QMME: 
$\dot{\hat{\rho}} = \mathcal{K}\hat{\rho}$, where $\hat{\rho}$ is the density matrix of S 
and $\mathcal{K}$ is the generator of the QMME (see Eq.~(\ref{QMME})).

Now suppose that the parameters are initially set to fixed values $\bm{\alpha}^{\rm i}$, 
as in Fig.~\ref{fig:example}(a). 
The system S is expected to be in a NESS characterized by $\bm{\alpha}^{\rm i}$. 
Then one changes $\bm{\alpha}$ along a curve $C$ that connects the initial values $\bm{\alpha}^{\rm i}$
and final values $\bm{\alpha}^{\rm f}$ in the parameter space. 
This operation is quasistatic 
if the time scale $\tau_{\rm op}$ of changing $\bm{\alpha}$ is sufficiently slow 
(more precisely, see the below of Eq.~(\ref{TDQMME})). 
In this case, S is expected to be in the instantaneous NESS at each time of the operation, 
and to settle finally to another NESS characterized by $\bm{\alpha}^{\rm f}$ (Fig.~\ref{fig:example}(b)).

One can measure (in principle) the entropy change in the reservoirs during an operation, 
which amounts to the entropy transferred from the reservoirs to the system S. 
This gives the entropy production in the system S during the operation. 
For quasistatic operations between NESSs, 
the total entropy production $\sigma_{\rm tot}$ diverges in time, 
since there is non-vanishing entropy flux at each time in the nonequilibrium case.
One of the ways of extracting the entropy production intrinsic to the operation 
is to use the excess entropy production, which is proposed in Ref.~\cite{OonoPaniconi}. 
Its rough definition is as follows (see Eq.~(\ref{excess}) for the precise definition).
For each $\bm{\alpha}$ in the operation path $C$, 
one can know the average entropy flux $J_\sigma$ for the instantaneous NESS determined by $\bm{\alpha}$.
By using this steady flux, one defines the excess entropy production as 
$\langle \sigma \rangle^{\rm ex} = \sigma_{\rm tot} - \int dt J_\sigma(\bm{\alpha}(t))$.

The main result of the present work is the geometrical expression for $\langle \sigma \rangle^{\rm ex}$, 
\begin{align}
\langle \sigma \rangle^{\rm ex} = \int_C \bm{A}(\bm{\alpha}) \cdot d\bm{\alpha}. 
\label{geometricalExcess}
\end{align}
The right-hand side is the Berry phase associated with the QMME generator $\mathcal{K}$, 
and $\bm{A}(\bm{\alpha})$ is the vector potential. 
This expression is a quantum extension of the result in Ref.~\cite{SagawaHayakawa} 
and valid for an arbitrary quasistatic operation between NESSs. 
Therefore, in general, the excess entropy production depends on the whole of the operation path $C$. 
This means the breakdown of the extended Clausius equality in the strongly nonequilibrium regime 
and suggests the importance of the vector potential $\bm{A}(\bm{\alpha})$ in the thermodynamics for NESSs.
It is also shown that this expression (\ref{geometricalExcess}) for $\langle \sigma \rangle^{\rm ex}$ 
reduces to Eq.~(\ref{equilibriumClausius}) in the equilibrium setups 
and to Eq.~(\ref{extendedClausius}) in the weakly nonequilibrium setups.

\subsection{Organization of this Paper}

This paper is organized as follows.
In Sec.~\ref{sec:QMME}, we explain the generic setup of the system which we consider in this paper.
We give the QMME description of the system of interest in the setup.
In Sec.~\ref{sec:FCS} we define the entropy production during an operation. 
We use the technique of the full counting statistics incorporated into the QMME \cite{EspositoHarbolaMukamel_2}
to calculate the cumulant generating function and average of the entropy production. 
In Sec.~\ref{sec:GQMME_explicit}, we derive an explicit form of the QMME by using the eigenoperators.
We also discuss the rotating wave approximation (or secular approximation) in the QMME.
We give the main result in Sec.~\ref{sec:main}; 
we derive the geometrical expression for the excess entropy production for an arbitrary quasistatic operation 
in the QMME system.
We also show that the results within and without the rotating wave approximation are equivalent. 
In Secs.~\ref{sec:equilibrium} and \ref{sec:weak}, 
we show that the geometrical expression reduces to the equilibrium and extended Clausius equalities 
(\ref{equilibriumClausius}) and (\ref{extendedClausius}) 
in the equilibrium states and in the weakly nonequilibrium regime, respectively.
In Sec.~\ref{sec:examples}, we investigate a spinless electron system in quantum dots as a simple example.
In Sec.~\ref{sec:summary}, we give a summary with a discussion.

\section{Setup}\label{sec:setup}

\subsection{Quantum Markovian Master Equation}\label{sec:QMME}

We consider a quantum system S in contact with reservoirs ${\rm R}_b$ ($b=1,2,...$). 
The system S and each reservoir can exchange particles and energy. 
We assume that the dimension of the Hilbert space $\mathcal{H}_{\rm S}$ associated with S is finite.
We also assume that the reservoirs are large enough compared to the system S.
The total system ${\rm S} + \{{\rm R}_b\}_b$ is closed except for external operations. 
Then the total system evolves according to the Liouville-von Neumann equation: 
\begin{align}
\frac{\partial \hat{\rho}_{\rm tot}(t)}{\partial t} 
= \frac{1}{i\hbar} [\hat{H}_{\rm tot} , \hat{\rho}_{\rm tot}(t)], 
\label{Liouville-vonNeumann}
\end{align}
where $\hat{\rho}_{\rm tot}$ and $\hat{H}_{\rm tot}$ denote respectively the density matrix 
and  Hamiltonian of the total system.
$\hat{H}_{\rm tot}$ is written as 
\begin{align}
\hat{H}_{\rm tot} = \hat{H}_{\rm S}(\bm{\alpha}_{\rm S}) + \sum_b \bigl[ \hat{H}_b + u \hat{H}_{{\rm S}b} \bigr], 
\label{Htot}
\end{align}
where $\hat{H}_{\rm S}$ is the system Hamiltonian, 
$\bm{\alpha}_{\rm S}$ is the set of operation parameters in the system S, 
$\hat{H}_b$ is the Hamiltonian of the $b$th reservoir ${\rm R}_b$, 
and $\hat{H}_{{\rm S}b}$ is the coupling Hamiltonian between S and ${\rm R}_b$. 
We assume 
that $\bigl[ \hat{H}_{\rm S}, \hat{N}_{\rm S} \bigr] = \bigl[ \hat{H}_b, \hat{N}_b \bigr] = 0$ holds 
with $\hat{N}_{\rm S}$ and $\hat{N}_b$ being the particle number operators of S and ${\rm R}_b$. 
We denote the eigenvalue of $\hat{H}_{\rm S}$ by $E_\nu$ and the corresponding eigenstate by $|E_\nu,n\rangle$, 
where $n$ is the index for distinguishing the degeneracy.
We also assume that the coupling between the system and reservoirs is weak. 
To keep in mind this weak coupling assumption, we introduced the parameter $u$ in Eq.~(\ref{Htot}). 

We set the initial states of the reservoirs to be equilibrium states 
with different temperatures and chemical potentials.
Precisely, the initial state of the reservoirs is 
$\hat{\rho}_{\rm R} = \bigotimes_b \hat{\rho}_b (\bm{\alpha}_b)$, 
where $\hat{\rho}_b(\bm{\alpha}_b) := e^{-\beta_b (\hat{H}_b - \mu_b \hat{N}_b)}/Z_b(\bm{\alpha}_b)$ 
is the grand canonical state of the $b$th reservoir 
with the inverse temperature $\beta_b$ and chemical potential $\mu_b$.
Here, $\bm{\alpha}_b$ denotes the set of the $b$th reservoir parameters ($\beta_b$, $\mu_b$), 
$Z_b(\bm{\alpha}_b) := {\rm Tr}_b e^{-\beta_b (\hat{H}_b - \mu_b \hat{N}_b)}$ is the grand partition function, 
and ${\rm Tr}_b$ is the trace over the degrees of freedom of the $b$th reservoir.
The initial state $\hat{\rho}(0)$ of the system S is arbitrary, 
so that the initial state of the total system is the uncorrelated state 
$\hat{\rho}_{\rm tot}(0) = \hat{\rho}(0) \otimes \hat{\rho}_{\rm R}$. 
Because the reservoirs are much larger than the system S, 
we expect that there exists a certain long time range 
in which the state of the system S can change considerably whereas the reservoirs approximately remain 
in the equilibrium states. 
We also expect that, in a more restricted but still long time range, the system settles down in a NESS 
which is uniquely determined by the reservoir parameters $\bm{\alpha}_{\rm B}:=\{\bm{\alpha}_b\}_b$ 
and system parameters $\bm{\alpha}_{\rm S}$.

We write the set of the control parameters $(\bm{\alpha}_{\rm S}, \bm{\alpha}_{\rm B})$ as $\bm{\alpha}$. 
An arbitrary external operation on the system S is represented by a modulation of $\bm{\alpha}$. 
Thus $\bm{\alpha}$ may depend on time.
We can treat the time-dependent $\bm{\alpha}$ within the framework 
of quantum Markovian master equation that we use in this work, 
as we will explain in and below Eq.~(\ref{TDQMME}). 

To investigate the dynamics of the system S in the above situation, 
we employ the quantum Markovian master equation (QMME) approach. 
In this approach the dynamics is described by an equation of motion for the reduced density matrix 
$\hat{\rho}={\rm Tr}_{\rm R} \hat{\rho}_{\rm tot}$ of S, 
where ${\rm Tr}_{\rm R}$ is the trace over the reservoirs. 
To derive the QMME we make the following assumption in addition to the weak coupling: 
the correlation time of the reservoirs is much shorter than the time scale of the system evolution. 
Starting from the Liouville-von Neumann equation (\ref{Liouville-vonNeumann}), 
and after tracing out the reservoirs' degrees of freedom, 
we perform the Born and Markov approximations on the basis of the above assumptions. 
For fixed $\bm{\alpha}$, the result is written in the Schr\"odinger picture as 
\cite{BreuerPetruccione,GardinerZoller}
\begin{align}
\frac{\partial \hat{\rho}(t)}{\partial t} 
= \frac{1}{i\hbar} \bigl[\hat{H}_{\rm S}(\bm{\alpha}_{\rm S}) , \hat{\rho}(t) \bigr]
- \frac{v}{\hbar^2} \sum_b \int_0^\infty dt' 
{\rm Tr}_b \Bigl[ \hat{H}_{{\rm S}b}, \bigl[ \Check{H}_{{\rm S}b}(-t'), 
\hat{\rho}(t) \otimes \hat{\rho}_b(\bm{\alpha}_b) \bigr] \Bigr], 
\label{QMME}
\end{align}
with $v:=u^2$. 
In this paper we refer to this equation as QMME. 
Here, the symbol `~$\Check{ }$~' stands for the Heisenberg picture 
in terms of $\hat{H}_{\rm S} + \sum_b  \hat{H}_b$, 
i.e., $\Check{O}(t) := \hat{U}^\dag(t) \hat{O} \hat{U}(t)$, 
where $\hat{U}(t) = \exp \bigl\{ - \bigl[ \hat{H}_{\rm S}(\bm{\alpha}_{\rm S}) 
+ \sum_b  \hat{H}_b \bigr] t /i\hbar\bigr\}$.
We note that the Born approximation is in the second order with respect to the system-reservoir coupling, 
which is represented by $v=u^2$ in the second term of Eq.~(\ref{QMME}).

We denote by $\mathsf{B}$ the set of all the linear operators on $\mathcal{H}_{\rm S}$. 
Because the dimension of $\mathcal{H}_{\rm S}$ is finite, 
any $\hat{Y}$ in $\mathsf{B}$ is a trace class operator.
We can define the Hilbert-Schmidt inner product in $\mathsf{B}$ as 
${\rm Tr}_{\rm S}(\hat{Y}_1^\dag \hat{Y}_2)$ for any $\hat{Y}_1, \hat{Y}_2 \in \mathsf{B}$, 
where ${\rm Tr}_{\rm S}$ is the trace in $\mathcal{H}_{\rm S}$. 
With this inner product, $\mathsf{B}$ is a separable Hilbert space. 
We refer to the linear operators on $\mathsf{B}$ as superoperators 
to distinguish with the operators on $\mathcal{H}_{\rm S}$. 
We define the adjoint $\mathcal{O}^\dag$ of a superoperator $\mathcal{O}$ such that  
${\rm Tr}_{\rm S} [(\mathcal{O}^\dag \hat{Y}_1)^\dag \hat{Y}_2] 
= {\rm Tr}_{\rm S} (\hat{Y}_1^\dag \mathcal{O} \hat{Y}_2)$ 
holds for any $\hat{Y}_1, \hat{Y}_2 \in \mathsf{B}$.

From the right-hand side (RHS) of the QMME (\ref{QMME}), 
we can define the generator (superoperator) $\mathcal{K}$ of the QMME as 
$\mathcal{K} \hat{Y} :=$ [RHS of Eq.~(\ref{QMME}) with $\hat{\rho}(t) \rightarrow \hat{Y}$] 
for any $\hat{Y} \in \mathsf{B}$. 
Since $\mathcal{K}$ depends on the control parameters $\bm{\alpha}$, 
we sometimes write them in the argument of the generator as $\mathcal{K}(\bm{\alpha})$. 

The right and left eigenvalue equations for $\mathcal{K}$ with fixed $\bm{\alpha}$ are respectively given by 
\begin{align}
\mathcal{K}(\bm{\alpha}) \hat{r}_m(\bm{\alpha}) &= \lambda_m(\bm{\alpha}) \hat{r}_m(\bm{\alpha}), 
\\
\mathcal{K}^\dag(\bm{\alpha}) \hat{\ell}_m(\bm{\alpha}) &= \lambda_m^*(\bm{\alpha}) \hat{\ell}_m(\bm{\alpha}), 
\end{align}
where the complex number $\lambda_m(\bm{\alpha})$ is the eigenvalue labeled by $m$ 
(we denote the complex conjugate of a complex number $c$ by $c^*$), 
and $\hat{r}_m(\bm{\alpha})$ and $\hat{\ell}_m(\bm{\alpha}) \in \mathsf{B}$ are respectively 
the corresponding right and left eigenvectors.
In the following, we assume that $\mathcal{K}(\bm{\alpha})$ has the zero eigenvalue $\lambda_0=0$ 
(labeled by $m=0$) without degeneracy, 
so that $\mathcal{K}(\bm{\alpha}) \hat{r}_0(\bm{\alpha}) = 0$ 
and $\mathcal{K}^\dag(\bm{\alpha}) \hat{\ell}_0(\bm{\alpha})=0$ hold.
This assumption implies that the QMME has a unique steady solution $\hat{\rho}_{\rm ss}(\bm{\alpha}) 
= \hat{r}_0(\bm{\alpha})$ for fixed $\bm{\alpha}$.
It should be noted, however, that the uniqueness of the steady solution of the QMME is not trivial 
especially in the case where $\hat{H}_{\rm S}$ has degenerate eigenenergies.
We note that $\hat{\ell}_0(\bm{\alpha}) = \hat{1}$ (identity operator on $\mathcal{H}_{\rm S}$) holds 
for any $\bm{\alpha}$ because of the trace-preserving property of the QMME.

When we modulate $\bm{\alpha}$ in time to perform a thermodynamic operation onto the system S, 
we can use the QMME with time-dependent $\bm{\alpha}$ for investigating the dynamics of S: 
\begin{align}
\frac{\partial \hat{\rho}(t)}{\partial t} = \mathcal{K}\bigl(\bm{\alpha}(t)\bigr) \hat{\rho}(t). 
\label{TDQMME}
\end{align}
This is valid for the operations whose time scale is sufficiently slower 
than the correlation time of the reservoirs.
This is a kind of Markov approximation other than the one used in deriving the QMME~(\ref{QMME}). 
There are four characteristic time scales in the present setup: 
the time scale $\tau_{\rm S}$ of the intrinsic evolution of the system S, 
the relaxation time  $\tau_{\rm rlx}$ of S as an open system in contact with the reservoirs, 
the correlation time $\tau_{\rm R}$ of the reservoirs, 
and the time scale $\tau_{\rm op}$ of the operation of changing $\bm{\alpha}$.
For the Markov approximation used here, $\tau_{\rm R} \ll \tau_{\rm op}$ is required, 
whereas $\tau_{\rm R} \ll \tau_{\rm rlx}$ is required for the Markov approximation in deriving Eq.~(\ref{QMME}). 
For the rotating wave approximation (or secular approximation), 
which will be explained in Sec.~\ref{sec:GQMME_explicit}, $\tau_{\rm S} \ll \tau_{\rm rlx}$ is required.
For quasistatic operations, required is $\tau_{\rm rlx} \ll \tau_{\rm op}$, 
which ensures the validity of an adiabatic approximation used in the next section. 

We here introduce the following projection superoperator $\mathcal{P}$: 
\begin{align}
\mathcal{P} |E_\kappa,k\rangle \langle E_\nu,n| = 
\begin{cases}
|E_\kappa,k\rangle \langle E_\nu,n| & (\text{if } E_\kappa = E_\nu)
\\
0 & (\text{if } E_\kappa \neq E_\nu).
\end{cases}
\end{align}
In the matrix representation of any operator $\hat{Y} \in \mathsf{B}$ 
in the basis of the eigenstates of $\hat{H}_{\rm S}$,
$\mathcal{P}$ leaves unchanged only the matrix elements constructed from the eigenstates 
with the same energy eigenvalues. 
By using $\mathcal{P}$, we define a subspace $\mathsf{P}$ of $\mathsf{B}$ as 
$\mathsf{P} := \{ \hat{Y} \in \mathsf{B} | ~ \mathcal{P}\hat{Y} = \hat{Y}\}$.
We denote the orthogonal complement of $\mathsf{P}$ by $\mathsf{Q}$ 
and the projection superoperator onto $\mathsf{Q}$ by $\mathcal{Q}$.

We also define the time-reversal operation. 
We denote the time-reversal operator on $\mathcal{H}_{\rm S}$ by $\hat{\theta}$. 
In this paper, we assume that the system Hamiltonian is time-reversal invariant: 
$\hat{\theta} \hat{H}_{\rm S} \hat{\theta}^{-1} = \hat{H}_{\rm S}$.
We also define the tilde superoperation on $\mathsf{B}$ by 
\begin{align}
\tilde{Y} := \hat{\theta} \hat{Y}^\dag \hat{\theta}^{-1}, 
\label{tilde}
\end{align}
for any $\hat{Y} \in \mathsf{B}$ \cite{Agarwal,Alicki}. 
We note that $\tilde{Y} = \hat{\theta} \hat{Y} \hat{\theta}^{-1}$ if $\hat{Y}$ is self-adjoint. 
Therefore the time reversal of a state $\hat{\rho}$ is given by 
$\hat{\theta} \hat{\rho} \hat{\theta}^{-1} = \tilde{\rho}$.
Using the superoperation (\ref{tilde}), 
we define the tilde $\tilde{\mathcal{O}}$ of a superoperator $\mathcal{O}$ by 
\begin{align}
\tilde{\mathcal{O}} \hat{Y} := \widetilde{\mathcal{O} \tilde{Y}}, 
\end{align}
for any $\hat{Y} \in \mathsf{B}$ \cite{Agarwal,Alicki}.

\subsection{Full Counting Statistics of Entropy Production}\label{sec:FCS}

We next introduce the entropy production $\sigma$ generated during an operation 
with a time interval $\tau$ as follows. 
At the initial time $t=0$, we perform a projection measurement of reservoir observables 
$\{\hat{A}_b(0) := \beta_b(0)(\hat{H}_b -\mu_b(0) \hat{N}_b)\}_b$ to obtain measurement outcomes $\{a_b(0)\}_b$.
Because we assume $[\hat{H}_b , \hat{N}_b] = 0$, 
the simultaneous projection measurement of $\hat{H}_b$ and $\hat{N}_b$ is possible. 
Note that we can construct the outcome $a_b(0)$ from the measurement outcomes of $\hat{H}_b$ and $\hat{N}_b$ 
because $\beta_b(0)$ and $\mu_b(0)$ are the initial (known) values of the control parameters.
For $t>0$, we make an operation by changing the control parameters $\bm{\alpha}$.
During the operation the system evolves with interacting with the reservoirs. 
At $t=\tau$, we again perform a projection measurement of 
$\{\hat{A}_b(\tau) := \beta_b(\tau)(\hat{H}_b -\mu_b(\tau) \hat{N}_b)\}_b$ 
to obtain measurement outcomes $\{a_b(\tau)\}_b$.
Since the reservoirs are assumed to remain in the equilibrium states during the operation, 
the difference of the outcomes gives the energy change $\sum_b \Delta E_b$ minus work $\sum_b \mu_b \Delta N_b$ 
associated with the particle exchange (divided by the temperature).
Therefore we can regard the difference of the outcomes as the heat (divided by the temperature) 
that is transferred from the reservoirs into the system S. 
We thus define the entropy production during the operation as 
\begin{align}
\sigma := \sum_b [a_b(0) - a_b(\tau)]. 
\label{entropyProduction}
\end{align}

By repeating this measurement scheme many times, 
we obtain a probability distribution $p_\tau(\sigma)$. 
We are interested in the average of $\sigma$, which is defined as 
$\langle \sigma \rangle_\tau := \int d\sigma p_\tau(\sigma) \sigma$. 
Note that we can define the steady value of the entropy flux $J_\sigma(\bm{\alpha})$ 
in a NESS by the asymptotic value of $\langle \sigma \rangle_\tau/\tau$: 
\begin{align}
J_\sigma(\bm{\alpha}) := \lim_{\tau\to\infty} \frac{\langle \sigma \rangle_\tau}{\tau}, 
\label{entropyFlux0}
\end{align}
with $\bm{\alpha}$ being fixed.

In this paper, instead of directly calculating the average, 
we investigate it from the cumulant generating function $G_\tau(\chi)$, which is given by 
\begin{align}
G_\tau(\chi) := \ln \int d\sigma p_\tau(\sigma) e^{i\chi\sigma}.
\end{align}
Here $\chi$ is the counting field. 
The derivatives of $G_\tau(\chi)$ give the cumulants; 
in particular, $\langle \sigma \rangle_\tau = \partial G_\tau(\chi) / \partial(i\chi)|_{\chi=0}$.

To calculate $G_\tau(\chi)$, we use a technique of the full counting statistics \cite{EspositoHarbolaMukamel_2}. 
This technique provides us with the formula for the cumulant generating function: 
$G_\tau(\chi) = \ln {\rm Tr} \hat{\rho}^\chi_{\rm tot}(\tau)$. 
Here, ${\rm Tr}$ is the trace over the total system 
and $\hat{\rho}^\chi_{\rm tot}$ is the solution of the generalized Liouville-von Neumann equation: 
\begin{align}
\frac{\partial \hat{\rho}^\chi_{\rm tot}(t)}{\partial t} 
= \frac{1}{i\hbar} \bigl[ \hat{H}_{\rm tot}^\chi \hat{\rho}^\chi_{\rm tot}(t) 
- \hat{\rho}^\chi_{\rm tot}(t) \hat{H}_{\rm tot}^{-\chi} \bigr].
\label{gLvN}
\end{align}
Here, the $\chi$-modified Hamiltonian $\hat{H}_{\rm tot}^\chi$ is given by 
\begin{align}
\hat{H}_{\rm tot}^\chi := e^{-i\chi \hat{A}/2} \hat{H}_{\rm tot} e^{i\chi \hat{A}/2}, 
\label{H_tot_chi}
\end{align}
where $\hat{A} := \sum_b \hat{A}_b$ with $\hat{A}_b := \beta_b(\hat{H}_b -\mu_b \hat{N}_b)$.

In the QMME approach, starting from the generalized Liouville-von Neumann equation (\ref{gLvN}), 
and taking the same procedure as in the previous subsection, 
we obtain the generalized quantum Markovian master equation (GQMME) 
for the reduced ($\chi$-modified) density matrix 
$\hat{\rho}^\chi = {\rm Tr}_{\rm R} \hat{\rho}_{\rm tot}^\chi$ as 
\begin{align}
\frac{\partial \hat{\rho}^{\chi}(t)}{\partial t} 
= & \frac{1}{i\hbar} \bigl[ \hat{H}_{\rm S}\bigl(\bm{\alpha}_{\rm S}\bigr) , \hat{\rho}^\chi(t) \bigr] 
- \frac{v}{\hbar^2} \sum_b \int_0^\infty dt' {\rm Tr}_b 
\Bigl[ \hat{H}_{{\rm S}b} , \bigl[ \Check{H}_{{\rm S}b}(-t') , 
\hat{\rho}^\chi(t) \otimes \hat{\rho}_b\bigl(\bm{\alpha}_b\bigr) \bigr]_\chi 
\Bigr]_\chi.
\label{GQMME}
\end{align}
Here, $[\hat{O}_1 , \hat{O}_2]_\chi 
:= \hat{O}_1^\chi \hat{O}_2 - \hat{O}_2 \hat{O}_1^{-\chi}$ 
and $\hat{O}^\chi := e^{-i\chi \hat{A}/2} \hat{O} e^{i\chi \hat{A}/2}$.
Thanks to the above-mentioned formula, 
we can calculate the generating function from the solution of the GQMME as 
\begin{align}
G_\tau(\chi) = \ln {\rm Tr}_{\rm S} \hat{\rho}^{\chi}(\tau).
\end{align}

Similarly to the case of the QMME, we can define the generator $\mathcal{K}^\chi$ of the GQMME as 
$\mathcal{K}^\chi(\bm{\alpha}) \hat{Y} :=$ [RHS of Eq.~(\ref{GQMME}) 
with $\hat{\rho}^\chi(t) \rightarrow \hat{Y}$] for any $\hat{Y} \in \mathsf{B}$. 
For fixed $\bm{\alpha}$, we can also define the right and left eigenvectors of 
the GQMME generator $\mathcal{K}^\chi(\bm{\alpha})$ 
corresponding to the eigenvalue $\lambda^\chi_m(\bm{\alpha})$, 
which are respectively denoted by $\hat{r}^\chi_m(\bm{\alpha})$ and $\hat{\ell}^\chi_m(\bm{\alpha})$.
They are normalized as ${\rm Tr}_{\rm S} (\hat{\ell}^{\chi\dag}_m \hat{r}^\chi_n) = \delta_{mn}$.
We assign the label for the eigenvalue with the maximum real part to $m=0$.
Then $\hat{\rho}^\chi(\tau) \sim e^{\lambda^\chi_0 \tau}$ holds for large $\tau$, 
which results in $\lim_{\tau\to\infty} G_\tau(\chi)/\tau = \lambda^\chi_0$.
Thus $\lambda^\chi_0(\bm{\alpha})$ is equal to the unit-time cumulant generating function $g(\chi)$
in the NESS for fixed $\bm{\alpha}$ \cite{EspositoHarbolaMukamel_2}.
Therefore, the average entropy flux $J_\sigma(\bm{\alpha})$ 
in the NESS, which is defined in Eq.~(\ref{entropyFlux0}), can be calculated by 
\begin{align}
J_\sigma(\bm{\alpha}) = \frac{\partial \lambda^\chi_0(\bm{\alpha})}{\partial (i\chi)}\bigg|_{\chi=0}.
\label{entropyFlux}
\end{align}
If we set $\chi=0$, the GQMME (\ref{GQMME}) reduces to the original QMME (\ref{QMME}), 
and $\mathcal{K}^\chi$, $\hat{\ell}^\chi_0$, and $\hat{r}^\chi_0$ also reduce to $\mathcal{K}$, $\hat{1}$, 
and $\hat{\rho}_{\rm ss}$, respectively. 
We also note that, for slow operations ($\tau_{\rm R} \ll \tau_{\rm op}$), 
we can use Eq.~(\ref{GQMME}) with time-dependent $\bm{\alpha}$: 
\begin{align}
\frac{\partial \hat{\rho}^{\chi}(t)}{\partial t} 
= \mathcal{K}^\chi\bigl(\bm{\alpha}(t)\bigr) \hat{\rho}^{\chi}(t). 
\label{GQMME2}
\end{align}

\subsection{Explicit Form of GQMME}\label{sec:GQMME_explicit}

We here consider the case where the system-reservoir couping Hamiltonian is given by 
\begin{align}
\hat{H}_{{\rm S}b} = \sum_l ( \hat{X}_{b,l} \otimes \hat{B}_{b,l}^\dag + \hat{X}_{b,l}^\dag \otimes \hat{B}_{b,l} ). 
\label{H_Sb_explicit}
\end{align}
Here, $\hat{X}_{b,l}$ and $\hat{B}_{b,l}$ ($\hat{X}_{b,l}^\dag$ and $\hat{B}_{b,l}^\dag$) 
are single-particle annihilation (creation) operators of the system S 
and of the $b$th reservoir ${\rm R}_b$, respectively, 
so that $[\hat{N}_{\rm S},\hat{X}_{b,l}] = -\hat{X}_{b,l}$, 
$[\hat{N}_{\rm S},\hat{X}_{b,l}^\dag] = \hat{X}_{b,l}^\dag$, 
$[\hat{N}_b,\hat{B}_{b,l}] = -\hat{B}_{b,l}$, 
and $[\hat{N}_b,\hat{B}_{b,l}^\dag] = \hat{B}_{b,l}^\dag$ hold. 
The index $l$ is a label for distinguishing the types of the coupling.
In this subsection, we derive an explicit form of the GQMME 
by introducing eigenoperators \cite{BreuerPetruccione,GardinerZoller}.

\subsubsection*{Eigenoperator}

Let $\hat{P}_b(\mathcal{E}_b)$ be the projection operator in ${\rm R}_b$ 
which projects onto the eigenspace belonging to the eigenvalue $\mathcal{E}_b$ of $\hat{H}_b$. 
Then we introduce the eigenoperators 
$\hat{B}_{b,l}^{(\Omega_b)}$ and $\hat{B}_{b,l}^{\dag(\Omega_b)}$ of $\hat{H}_b$ as 
\begin{align}
\hat{B}_{b,l}^{(\Omega_b)} &:= \sum_{\mathcal{E}_b} 
\hat{P}_b(\mathcal{E}_b - \hbar\Omega_b) \hat{B}_{b,l} \hat{P}_b(\mathcal{E}_b), 
\\
\hat{B}_{b,l}^{\dag(\Omega_b)} &:= \sum_{\mathcal{E}_b} 
\hat{P}_b(\mathcal{E}_b + \hbar\Omega_b) \hat{B}_{b,l}^\dag \hat{P}_b(\mathcal{E}_b), 
\end{align}
where $\hbar\Omega_b$ is a difference of the reservoir eigenenergies.
$\hat{B}_{b,l}^{(\Omega_b)}$ ($\hat{B}_{b,l}^{\dag(\Omega_b)}$) 
decreases (increases) the energy and particle number 
of the reservoir ${\rm R}_b$ by $\hbar\Omega_b$ and 1, respectively.
We note that $\hat{B}_{b,l}$ and $\hat{B}_{b,l}^\dag$ can be decomposed into the eigenoperators: 
\begin{align}
\hat{B}_{b,l} &= \sum_{\Omega_b} \hat{B}_{b,l}^{(\Omega_b)}, 
\label{B_bl}
\\
\hat{B}_{b,l}^\dag &= \sum_{\Omega_b} \hat{B}_{b,l}^{\dag(\Omega_b)}. 
\label{B_bl^dag}
\end{align}

Similarly, we introduce the eigenoperators of the system S.
Let $\hat{P}_{\rm S}(E_\nu)$ be the projection operator in S 
which projects onto the eigenspace belonging to the eigenvalue $E_\nu$ of $\hat{H}_{\rm S}$.
Then we define the eigenoperators 
$\hat{X}_{b,l}^{(\omega_{\rm S})}$ and $\hat{X}_{b,l}^{\dag (\omega_{\rm S})}$ 
of $\hat{H}_{\rm S}$ as 
\begin{align}
\hat{X}_{b,l}^{(\omega_{\rm S})} 
&:= \sum_\nu \hat{P}_{\rm S} (E_\nu - \hbar\omega_{\rm S}) \hat{X}_{b,l} \hat{P}_{\rm S} (E_\nu), 
\\
\hat{X}_{b,l}^{\dag (\omega_{\rm S})} 
&:= \sum_\nu \hat{P}_{\rm S} (E_\nu + \hbar\omega_{\rm S}) \hat{X}_{b,l}^\dag \hat{P}_{\rm S} (E_\nu), 
\end{align}
where $\hbar\omega_{\rm S}$ is a difference of the system eigenenergies.
$\hat{X}_{b,l}^{(\omega_{\rm S})}$ ($\hat{X}_{b,l}^{\dag(\omega_{\rm S})}$) decreases (increases) 
the energy and particle number of the system S by $\hbar\omega_{\rm S}$ and 1, respectively.
We note that $\hat{X}_{b,l}$ and $\hat{X}_{b,l}^\dag$ are reconstructed from the eigenoperators: 
\begin{align}
\hat{X}_{b,l} &= \sum_{\omega_{\rm S}} \hat{X}_{b,l}^{(\omega_{\rm S})}, 
\label{X_bl}
\\
\hat{X}_{b,l}^\dag &= \sum_{\omega_{\rm S}} \hat{X}_{b,l}^{\dag(\omega_{\rm S})}. 
\label{X_bl^dag}
\end{align}

\subsubsection*{GQMME Generator with Eigenoperators}

By substituting Eqs.~(\ref{H_Sb_explicit}), (\ref{B_bl}), (\ref{B_bl^dag}), (\ref{X_bl}), and (\ref{X_bl^dag})
into Eq.~(\ref{GQMME}), and using the properties of the eigenoperators, 
we obtain an explicit form of the GQMME generator as 
\begin{align}
\mathcal{K}^\chi = \mathcal{K}_0 + v \sum_b \mathcal{L}_b^\chi, 
\label{GQMME_explicit}
\end{align}
where 
\begin{align}
\mathcal{K}_0 \hat{Y} &= \frac{1}{i\hbar} \bigl[ \hat{H}_{\rm S} , \hat{Y} \bigr],
\label{GQMME_explicit0}
\\
\mathcal{L}_b^\chi \hat{Y} &= - \frac{1}{2\hbar^2} \sum_{l,l'} \sum_{\omega_{\rm S},\omega_{\rm S}'}
\biggl[ 
\Phi_{b,ll'}^+ (\omega_{\rm S}') \Bigl\{ 
\hat{X}_{b,l}^{(\omega_{\rm S})} \hat{X}_{b,l'}^{\dag(\omega_{\rm S}')} \hat{Y} 
+ \hat{Y} \hat{X}_{b,l}^{(\omega_{\rm S}')} \hat{X}_{b,l'}^{\dag(\omega_{\rm S})} 
\notag\\
&\hspace{48mm}
- e^{i \chi \beta_b (\hbar\omega_{\rm S}' - \mu_b)} \Bigl( 
\hat{X}_{b,l'}^{\dag(\omega_{\rm S})} \hat{Y} \hat{X}_{b,l}^{(\omega_{\rm S}')} 
+ \hat{X}_{b,l'}^{\dag(\omega_{\rm S}')} \hat{Y} \hat{X}_{b,l}^{(\omega_{\rm S})} \Bigr) \Bigr\} 
\notag\\
&\hspace{27mm}
+ \Phi_{b,ll'}^- (\omega_{\rm S}') \Bigl\{ 
\hat{X}_{b,l}^{\dag(\omega_{\rm S})} \hat{X}_{b,l'}^{(\omega_{\rm S}')} \hat{Y} 
+ \hat{Y} \hat{X}_{b,l}^{\dag(\omega_{\rm S}')} \hat{X}_{b,l'}^{(\omega_{\rm S})} 
\notag\\
&\hspace{48mm}
- e^{-i \chi \beta_b (\hbar\omega_{\rm S}' - \mu_b)} \Bigl( 
\hat{X}_{b,l'}^{(\omega_{\rm S})} \hat{Y} \hat{X}_{b,l}^{\dag(\omega_{\rm S}')} 
+ \hat{X}_{b,l'}^{(\omega_{\rm S}')} \hat{Y} \hat{X}_{b,l}^{\dag(\omega_{\rm S})} \Bigr) 
\Bigr\} \biggr], 
\label{GQMME_explicit1}
\end{align}
Here $\Phi_{b,ll'}^\pm (\omega)$ is referred to as spectral function of the $b$th reservoir, 
which is given by 
\begin{align}
\Phi_{b,ll'}^+ (\omega) 
&:= 2\pi \sum_{\Omega_b} \delta(\Omega_b - \omega) {\rm Tr}_b \Bigl\{ \hat{\rho}_b(\bm{\alpha}_b) 
\hat{B}_{b,l}^{\dag(\Omega_b)} \hat{B}_{b,l'}^{(\Omega_{b})} \Bigr\}, 
\label{Phi+}
\\
\Phi_{b,ll'}^- (\omega) 
&:= 2\pi \sum_{\Omega_b} \delta(\Omega_b - \omega) {\rm Tr}_b \Bigl\{ \hat{\rho}_b(\bm{\alpha}_b) 
\hat{B}_{b,l}^{(\Omega_b)} \hat{B}_{b,l'}^{\dag(\Omega_{b})} \Bigr\}. 
\label{Phi-}
\end{align}
In Eq.~(\ref{GQMME_explicit}), we neglected the terms which are proportional to the Hilbert transform of 
$\Phi_{b,ll'}^\pm (\omega)$ because these terms are known to give negligible contribution to the dynamics
\cite{GaspardNagaoka}.
The spectral function has the following properties: 
\begin{align}
\bigl[ \Phi_{b,ll'}^+ (\omega) \bigr]^* &= \Phi_{b,l'l}^+ (\omega), 
\\
\Phi_{b,ll'}^+ (\omega) &= e^{-\beta_b (\hbar\omega - \mu_b)} \Phi_{b,l'l}^- (\omega). 
\label{KMS}
\end{align}
The latter is the Kubo-Martin-Schwinger (KMS) condition.

For later use, we introduce the $i\chi$-derivative of the generator: 
$\mathcal{K}^\prime := \partial \mathcal{K}^\chi / \partial (i\chi)|_{\chi=0}$.
From Eq.~(\ref{GQMME_explicit}), we can write $\mathcal{K}'$ as 
$\mathcal{K}' = v \sum_b \mathcal{L}'_b$
with $\mathcal{L}_b^\prime := \partial \mathcal{L}_b^\chi / \partial (i\chi)|_{\chi=0}$. 
Moreover, we can show that 
\begin{align}
\mathcal{P} \mathcal{L}_b^\prime \mathcal{P} \hat{Y} 
= \beta_b \bigl( \hat{H}_{\rm S} - \mu_b \hat{N}_{\rm S} \bigr) \mathcal{P} \mathcal{L}_b \mathcal{P} \hat{Y} 
- \mathcal{P} \mathcal{L}_b \mathcal{P} \beta_b \bigl( \hat{H}_{\rm S} - \mu_b \hat{N}_{\rm S} \bigr) \hat{Y} 
\label{L_b_prime_L_b}
\end{align}
holds for any $\hat{Y} \in \mathsf{B}$, where $\mathcal{L}_b = \mathcal{L}_b^{\chi=0}$.

\subsubsection*{Rotating Wave Approximation}

We here consider the situation where the time scale $\tau_{\rm S}$ of the intrinsic evolution of the system S 
is much smaller than the relaxation time $\tau_{\rm rlx}$ of S, 
where $\tau_{\rm S}$ is given by a typical value of $|\omega_{\rm S} - \omega_{\rm S}'|^{-1}$ 
and $\tau_{\rm rlx}$ is the time scale over which S varies appreciably.
In this case, the terms with $\omega_{\rm S} \neq \omega_{\rm S}'$ in Eq.~(\ref{GQMME_explicit1}) 
rapidly oscillate within the time scale $\tau_{\rm rlx}$ if they are written in the interaction picture.
Therefore we may neglect these terms and leave only the terms with $\omega_{\rm S} = \omega_{\rm S}'$. 
This approximation is known as the rotating wave approximation (RWA) or secular approximation \cite{BreuerPetruccione}.
We express the quantities and (super)operators within the RWA by the subscript `r'; 
for example, we write the GQMME generator within the RWA as $\mathcal{K}^\chi_{\rm r}$. 

Similarly to the case without the RWA, we can decompose the generator $\mathcal{K}^\chi_{\rm r}$ as 
\begin{align}
\mathcal{K}_{\rm r}^\chi 
= \mathcal{K}_0 + v \sum_b \mathcal{L}_{b,\rm r}^\chi, 
\label{K_r_chi}
\end{align}
where we define $\mathcal{L}_{b,\rm r}^\chi$ by leaving only the 
$\omega_{\rm S}' = \omega_{\rm S}$ terms in the $\omega_{\rm S}'$-sum in Eqs.~(\ref{GQMME_explicit1}).
We can show the following equation 
\begin{align}
\mathcal{Q} \mathcal{K}_{\rm r}^\chi \mathcal{P} = \mathcal{P} \mathcal{K}_{\rm r}^\chi \mathcal{Q} =0.
\label{coherencePartIs0_RWA}
\end{align}
This implies that the GQMME is decomposed into two closed systems of equations: 
one is for $\mathcal{P}\hat{\rho}^\chi_{\rm r}$ and the other is for $\mathcal{Q}\hat{\rho}^\chi_{\rm r}$.
Furthermore, we can show that $\mathcal{P} \mathcal{K}^\chi_{\rm r} \mathcal{P} 
= \mathcal{P} \mathcal{K}^\chi \mathcal{P}$ holds. 

We write the right and left eigenvectors of $\mathcal{K}^\chi_{\rm r}$ 
corresponding to the eigenvalue $\lambda_{m,\rm r}^\chi$ 
as $\hat{r}_{m,\rm r}^\chi$ and $\hat{\ell}_{m,\rm r}^\chi$. 
They are normalized as 
${\rm Tr}_{\rm S} (\hat{\ell}^{\chi\dag}_{m,\rm r} \hat{r}^\chi_{n,\rm r}) = \delta_{mn}$.
We assign the label for the eigenvalue with the maximum real part to $m=0$. 
For the reason mentioned below Eq.~(\ref{coherencePartIs0_RWA}), 
we can classify the eigenvectors into two groups: 
one group is in $\mathsf{P}$ and the other is in $\mathsf{Q}$.
In particular, the eigenvectors for $m=0$ belong to the former group. 
We note that $\hat{\ell}_{0,\rm r}^{\chi=0} = \hat{1}$ holds, 
and that $\hat{\rho}_{\rm ss,r} := \hat{r}_{0,\rm r}^{\chi=0} $ 
is the steady solution of the QMME within the RWA.
We assume that the steady solution is uniquely determined for fixed $\bm{\alpha}$ also in the RWA.

\subsubsection*{Remark on the fluctuation theorem}

Before closing this section, we make a remark on the fluctuation theorem (FT) in the present setup.
First, we consider in the RWA.
Due to the KMS condition (\ref{KMS}), $\mathcal{L}_{b,\rm r}^\chi$ in Eq.~(\ref{K_r_chi}) satisfies 
$\mathcal{L}_{b,\rm r}^{-\chi-i} = \mathcal{L}_{b,\rm r}^{-\chi\dag}$. 
Moreover, because $\hat{\ell}_{0,\rm r}^\chi$ is in $\mathsf{P}$, 
$\mathcal{K}_0 \hat{\ell}_{0,\rm r}^\chi = \mathcal{K}_0^\dag \hat{\ell}_{0,\rm r}^\chi = 0$ holds.
Therefore, we have $\mathcal{K}_{\rm r}^{-\chi-i} \hat{\ell}_{0,\rm r}^{-\chi}
= \mathcal{K}_{\rm r}^{-\chi\dag} \hat{\ell}_{0,\rm r}^{-\chi} 
= \lambda_{0,\rm r}^{-\chi*} \hat{\ell}_{0,\rm r}^{-\chi}$; 
i.e., the eigenvalue $\lambda_{0,\rm r}^{-\chi-i}$ of $\mathcal{K}_{\rm r}^{-\chi-i}$ with the maximum real part 
is equal to $\lambda_{0,\rm r}^{-\chi*}$.
Next, we consider without the RWA. 
In Sec.~\ref{sec:main} we will show that $\lambda_0^{\chi} = \lambda_{0,\rm r}^{\chi} + O(v^2)$ holds. 
Therefore we obtain $\lambda_0^{-\chi-i} = \lambda_0^{-\chi*} + O(v^2)$.
Note that $O(v)$ is the same precision as that in the QMME.

This equality leads to the relation $g(-\chi -i)=g^*(-\chi) + O(v^2)$, 
because $\lambda_0^\chi$ is equal to the unit-time cumulant generating function $g(\chi)$ 
of the entropy production in the NESS for fixed $\bm{\alpha}$. 
Noting that the generating function satisfies $g(-\chi)=g^*(\chi)$, 
we obtain the symmetry relation $g(-\chi -i)=g(\chi) + O(v^2)$. 
This is one of the expressions of the steady-state FT for the entropy production
\cite{EspositoHarbolaMukamel_2}.
This result supports that our definition (\ref{entropyProduction}) of the entropy production is reasonable.
Note that the fluctuation theorem shown here is the steady state FT, 
where the control parameters $\bm{\alpha}$ are fixed. 
On the other hand, if a modulation of $\bm{\alpha}$ is considered as in the next section, 
the transient FT should hold (although we do not show in this paper).

\section{Results for Generic System}\label{sec:results}

\subsection{Geometrical Expression of Excess Entropy Production}\label{sec:main}

We now consider an arbitrary quasistatic operation that connects two steady states.
At the initial time $t=0$, the system S is set to be in a steady state that is uniquely specified by $\bm{\alpha}(0)=\bm{\alpha}^{\rm i}$. 
Then the system S is subjected to a thermodynamic operation that is characterized by a modulation of 
$\bm{\alpha}$ along a curve $C$ in the parameter space. 
At $t=\tau$, S settles in another steady state with $\bm{\alpha}(\tau)=\bm{\alpha}^{\rm f}$.
For the quasistatic operation, the time interval $\tau$ of the operation is sufficiently larger 
than the relaxation time scale of the system. 
Since there exist steady particle and energy currents in the NESS at each $\bm{\alpha}$ in $C$, 
the average entropy production $\langle \sigma \rangle_\tau$ includes a component 
that linearly increases with $\tau$. 
This component is referred to as house-keeping part of the entropy production 
\cite{OonoPaniconi} and is given by 
\begin{align}
\langle \sigma \rangle_\tau^{\rm hk} := \int_0^\tau dt J_\sigma \bigl(\bm{\alpha}(t) \bigr),
\label{houseKeeping}
\end{align}
where $J_\sigma(\bm{\alpha})$ is the steady entropy flux defined in Eq.~(\ref{entropyFlux0}) 
for fixed $\bm{\alpha}$.
By subtracting this component from $\langle \sigma \rangle_\tau$, we define the excess entropy production: 
\begin{align}
\langle \sigma \rangle^{\rm ex} := \langle \sigma \rangle_\tau - \langle \sigma \rangle_\tau^{\rm hk}.
\label{excess}
\end{align}
As we shall show in the below, $\langle \sigma \rangle^{\rm ex}$ is independent of $\tau$ 
for the quasistatic operation. 

The main result of this paper is the geometrical expression~(\ref{geometricalExcess}) 
for $\langle \sigma \rangle^{\rm ex}$, 
where the vector potential is identified as 
\begin{align}
\bm{A}(\bm{\alpha}) &:= - {\rm Tr}_{\rm S} \left[ \hat{\ell}^{\prime\dag}_0(\bm{\alpha}) 
\frac{\partial \hat{\rho}_{\rm ss}(\bm{\alpha})}{\partial \bm{\alpha}} \right], 
\label{vectorPotential}
\end{align}
with $\hat{\ell}_0^{\prime} := \partial \hat{\ell}_0^\chi/\partial(i\chi)|_{\chi=0}$. 
We will give the derivation later in this subsection. 
This expression holds for any quasistatic operations between arbitrary NESSs 
if the system is described by the QMME.
The RHS of Eq.~(\ref{geometricalExcess}) is analogous to the Berry phase in quantum mechanics \cite{Berry}.
It is geometrical because it depends only on the line integral along the curve $C$ but not on $\tau$.
This implies that in general 
the excess entropy production is not given by the difference of a scalar potential. 
In the equilibrium states and the weakly nonequilibrium regime, 
the vector potential $\bm{A}$ is given by the gradient of a scalar potential, 
as will be shown in the following subsections, 
so that the excess entropy production is written as the difference of it. 
In the strongly nonequilibrium regime, in contrast, 
it seems that $\bm{A}$ is not a gradient in most systems and operations except for special cases, 
such as an example given in Sec.~\ref{sec:examples}. 

Furthermore, as we will show later in this subsection, 
the analyses within and without the RWA give the equivalent result 
for the geometrical expression (\ref{geometricalExcess}).
That is, the following relation holds for the vector potential: 
\begin{align}
\bm{A}(\bm{\alpha}) = \bm{A}_{\rm r}(\bm{\alpha}) + O(v^2), 
\label{integrand}
\end{align}
where 
\begin{align}
\bm{A}_{\rm r}(\bm{\alpha}) 
:= - {\rm Tr}_{\rm S} \left[ \hat{\ell}^{\prime\dag}_{0,\rm r}(\bm{\alpha}) 
\frac{\partial \hat{\rho}_{\rm ss,r}(\bm{\alpha})}{\partial \bm{\alpha}} \right], 
\label{vectorPotentialRWA}
\end{align}
and $\hat{\ell}_{0,\rm r}^{\prime} := \partial \hat{\ell}_{0,\rm r}^\chi/\partial(i\chi)|_{\chi=0}$.
Because the QMME is valid up to $O(v)$, this equation implies the equivalence 
between the vector potentials within and without the RWA.
Therefore we can safely use the RWA to calculate the excess entropy production 
$\langle \sigma \rangle^{\rm ex}$, 
whereas it is known that the internal current in the system S vanishes in NESSs under the RWA 
\cite{Wichterich_etal}.

\subsubsection*{Derivation of Eq.~(\ref{geometricalExcess})}

We first note that the excess entropy production can be written as 
$\langle \sigma \rangle^{\rm ex} = \partial G^{\rm ex}(\chi) / \partial(i\chi)|_{\chi=0}$, where 
\begin{align}
G^{\rm ex}(\chi) := G_\tau(\chi) - \int_0^\tau dt \lambda^\chi_0 \bigl(\bm{\alpha}(t) \bigr)
\end{align}
is the excess part of the cumulant generating function of the entropy production.
This is because $\langle \sigma \rangle_\tau = \partial G_\tau(\chi) / \partial(i\chi)|_{\chi=0}$ 
and $J_\sigma(\bm{\alpha})= \partial \lambda^\chi_0(\bm{\alpha})/\partial (i\chi)|_{\chi=0}$ 
as is mentioned in the previous section. 

We derive the geometrical expression for $G^{\rm ex}(\chi)$ by using the method 
similar to those in Refs~\cite{SagawaHayakawa,SinitsynNemenman2007a,Yuge_etal}. 
To solve the GQMME for a given curve $C$ of $\bm{\alpha}$ in the parameter space, 
we expand $\hat{\rho}^\chi(t)$ as 
\begin{align}
\hat{\rho}^\chi(t) = \sum_m c_m(t) e^{\Lambda^\chi_m(t)} \hat{r}^\chi_m \bigl(\bm{\alpha}(t)\bigr),
\label{expansion_rho}
\end{align}
where $\Lambda^\chi_m(t) := \int_0^t dt' \lambda^\chi_m \bigl(\bm{\alpha}(t')\bigr)$.
Substituting this expansion into Eq.~(\ref{GQMME2}),
we rewrite the left-hand side (LHS) and the RHS of Eq.~(\ref{GQMME2}) respectively as 
\begin{align*}
[\text{LHS of Eq.} (\ref{GQMME2})] 
&= \sum_m e^{\Lambda^\chi_m(t)} \Bigl[ \dot{c}_m(t) \hat{r}^\chi_m \bigl(\bm{\alpha}(t)\bigr) 
+ \lambda^\chi_m \bigl(\bm{\alpha}(t)\bigr) c_m(t) \hat{r}^\chi_m \bigl(\bm{\alpha}(t)\bigr) 
+ c_m(t) \dot{\hat{r}}^\chi_m \bigl(\bm{\alpha}(t)\bigr) \Bigr],
\\
[\text{RHS of Eq.} (\ref{GQMME2})] 
&= \sum_m e^{\Lambda^\chi_m(t)} \lambda^\chi_m \bigl(\bm{\alpha}(t)\bigr) c_m(t) 
\hat{r}^\chi_m \bigl(\bm{\alpha}(t)\bigr),
\end{align*}
where `~$\dot{}$~' stands for the time derivative.
Equating these equations and taking the Hilbert-Schmidt inner product 
with $\hat{\ell}^\chi_0 \bigl(\bm{\alpha}(t)\bigr)$, we obtain 
\begin{align}
\dot{c}_0(t) 
= - \sum_m c_m(t) e^{\Lambda^\chi_m(t)-\Lambda^\chi_0(t)} 
{\rm Tr}_{\rm S} \left[ \hat{\ell}^{\chi\dag}_0 \bigl(\bm{\alpha}(t)\bigr) 
\dot{\hat{r}}^\chi_m \bigl(\bm{\alpha}(t)\bigr) \right].
\label{eq_c0}
\end{align}
If the time scale of the modulation of $\bm{\alpha}$ is much slower than that of the relaxation of the system, 
we can approximate the sum on the RHS of Eq.~(\ref{eq_c0}) by the contribution only from the term with $m=0$ 
(note that ${\rm Re} [\Lambda^\chi_m(t)-\Lambda^\chi_0(t)] < 0$ for $m \neq 0$):
\begin{align}
\dot{c}_0(t) 
\simeq - c_0(t) {\rm Tr}_{\rm S} \left[ \hat{\ell}^{\chi\dag}_0 \bigl(\bm{\alpha}(t)\bigr) 
\dot{\hat{r}}^\chi_0 \bigl(\bm{\alpha}(t)\bigr) \right].
\end{align}
This approximation corresponds to the adiabatic approximation in quantum mechanics.
By solving this approximate equation we obtain 
\begin{align}
c_0 (\tau) 
= c_0(0) \exp\left\{ - \int_{C} {\rm Tr}_{\rm S} \left[ \hat{\ell}^{\chi\dag}_0(\bm{\alpha}) 
d \hat{r}^\chi_0(\bm{\alpha}) \right] \right\}, 
\label{c_0}
\end{align}
where $d \hat{r}^\chi_0(\bm{\alpha}) 
:= \bigl(\partial \hat{r}^\chi_0(\bm{\alpha}) / \partial \bm{\alpha} \bigr) \cdot d \bm{\alpha}$.
If the initial state of the system S is 
$\hat{\rho}^\chi(0) = \hat{\rho}_{\rm ss} \bigl(\bm{\alpha}^{\rm i}\bigr)$, 
then $c_0(0) = {\rm Tr}_{\rm S}\Bigl[ \hat{\ell}^{\chi\dag}_0\bigl(\bm{\alpha}^{\rm i}\bigr) 
\hat{\rho}_{\rm ss}\bigl(\bm{\alpha}^{\rm i}\bigr) \Bigr]$.
We substitute Eq.~(\ref{c_0}) into the $m=0$ term in Eq.~(\ref{expansion_rho}).
At long time only the $m=0$ term remains and $m \neq 0$ terms vanish in Eq.~(\ref{expansion_rho}) 
since $\Lambda^\chi_0(t)$ has the maximum real part. 
Therefore we obtain 
\begin{align}
\hat{\rho}^\chi(\tau) &\simeq c_0 (\tau) e^{\Lambda^\chi_0(\tau)} \hat{r}^\chi_0 \bigl(\bm{\alpha}^{\rm f}\bigr)
\notag\\
&= e^{\Lambda^\chi_0(\tau)} \hat{r}^\chi_0 \bigl(\bm{\alpha}^{\rm f}\bigr) 
{\rm Tr}_{\rm S} \Bigl[ \hat{\ell}^{\chi\dag}_0 \bigl(\bm{\alpha}^{\rm i}\bigr) 
\hat{\rho}_{\rm ss} \bigl(\bm{\alpha}^{\rm i}\bigr) \Bigr] 
\exp\left\{ - \int_{C} {\rm Tr}_{\rm S} \bigl[ \hat{\ell}^{\chi\dag}_0(\bm{\alpha}) 
d \hat{r}^\chi_0(\bm{\alpha}) \bigr] \right\}. 
\end{align}
We thus obtain the excess cumulant generating function 
$G^{\rm ex}(\chi) = \ln {\rm Tr}_{\rm S} \hat{\rho}^\chi(\tau) - \Lambda^\chi_0(\tau)$ for the slow modulation: 
\begin{align}
G^{\rm ex}(\chi) &= - \int_C {\rm Tr}_{\rm S} \Bigl[ \hat{\ell}_0^{\chi\dag}(\bm{\alpha}) d \hat{r}_0^\chi(\bm{\alpha}) \Bigr] 
+ \ln {\rm Tr}_{\rm S} \Bigl[ \hat{\ell}_0^{\chi\dag} \bigl(\bm{\alpha}^{\rm i}\bigr) \hat{\rho}_{\rm ss} \bigl(\bm{\alpha}^{\rm i}\bigr) \Bigr] 
+ \ln {\rm Tr}_{\rm S} \hat{r}_0^{\chi} \bigl(\bm{\alpha}^{\rm f}\bigr) . 
\label{CGF_ex}
\end{align}
Equation~(\ref{CGF_ex}) is analogous to the Berry phase, 
and $\Lambda^\chi_0(\tau)$ corresponds to the dynamical phase.

By differentiating Eq.~(\ref{CGF_ex}) with respect to $i\chi$ and setting $\chi=0$, 
we obtain the expression for the excess entropy production: 
\begin{align}
\langle \sigma \rangle^{\rm ex} 
&= - \int_C {\rm Tr}_{\rm S}\left[ \hat{\ell}_0^{\prime\dag}(\bm{\alpha}) d\hat{r}_0^\chi(\bm{\alpha}) \right] 
- \int_C {\rm Tr}_{\rm S}\left[ d\hat{r}_0^\prime(\bm{\alpha}) \right] 
+ {\rm Tr}_{\rm S} \Bigl[ \hat{\ell}_0^{\prime\dag}\bigl(\bm{\alpha}^{\rm i}\bigr) \hat{\rho}_{\rm ss} \bigl(\bm{\alpha}^{\rm i}\bigr) \Bigr] 
+ {\rm Tr}_{\rm S} \hat{r}_0^{\prime} \bigl(\bm{\alpha}^{\rm f}\bigr)
\notag\\
&= - \int_C {\rm Tr}_{\rm S}\left[ \hat{\ell}_0^{\prime\dag}(\bm{\alpha}) d\hat{r}_0^\chi(\bm{\alpha}) \right] 
+ {\rm Tr}_{\rm S}\left[ \hat{r}_0^\prime\bigl(\bm{\alpha}^{\rm i}\bigr) \right] 
+ {\rm Tr}_{\rm S}\left[ \hat{\ell}_0^{\prime\dag}\bigl(\bm{\alpha}^{\rm i}\bigr) \hat{\rho}_{\rm ss} \bigl(\bm{\alpha}^{\rm i}\bigr) \right] 
\notag\\
&= - \int_C {\rm Tr}_{\rm S}\left[ \hat{\ell}_0^{\prime\dag}(\bm{\alpha}) d\hat{r}_0^\chi(\bm{\alpha}) \right]. 
\label{excess_derived}
\end{align}
We thus obtain Eq.~(\ref{geometricalExcess}).
In Eq.~(\ref{excess_derived}), we introduced 
$\hat{r}_0^\prime := \partial \hat{r}_0^\chi/\partial (i\chi) |_{\chi=0}$, 
and used $\hat{\ell}_0 = \hat{1}$ in the first line.
In the third line, the surface terms vanish because they are rewritten as 
\begin{align}
{\rm Tr}_{\rm S}\left[ \hat{r}_0^\prime\bigl(\bm{\alpha}^{\rm i}\bigr) \right] 
+ {\rm Tr}_{\rm S}\left[ \hat{\ell}_0^{\prime\dag}\bigl(\bm{\alpha}^{\rm i}\bigr) \hat{\rho}_{\rm ss} \bigl(\bm{\alpha}^{\rm i}\bigr) \right] 
= \frac{\partial}{\partial (i\chi)} {\rm Tr}_{\rm S}\left[ \hat{\ell}_0^{\chi\dag}\bigl(\bm{\alpha}^{\rm i}\bigr) \hat{r}_0^\chi \bigl(\bm{\alpha}^{\rm i}\bigr) \right] \bigg|_{\chi=0}, 
\end{align}
and because of the normalization condition 
${\rm Tr}_{\rm S}\left[ \hat{\ell}_0^{\chi\dag}\bigl(\bm{\alpha}^{\rm i}\bigr) 
\hat{r}_0^\chi \bigl(\bm{\alpha}^{\rm i}\bigr) \right] = 1$.

\subsubsection*{Derivation of Eq. (\ref{integrand})}

To show the equivalence between the results within and without the RWA, 
we derive the relations between the eigenvalues and eigenvectors of the GQMME generators 
within and without the RWA.
For this purpose, we decompose the generator 
as $\mathcal{K}^\chi = \mathcal{K}^\chi_{\rm r} + v \mathcal{R}^\chi$. 
Then $v \mathcal{R}^\chi$ is $O(v)$ since $\mathcal{K}^\chi_{\rm r}$ 
contains all of the $O(v^0)$ terms (as well as a part of the $O(v)$ terms) in $\mathcal{K}^\chi$. 
Motivated by this decomposition, we here make the ansatz that 
we can expand the eigenvalue $\lambda_0^\chi$ and eigenvectors $\hat{\ell}_0^\chi$ and $\hat{r}_0^\chi$ 
of $\mathcal{K}^\chi$ around those of $\mathcal{K}^\chi_{\rm r}$ with respect to $v$: 
\begin{align}
\lambda_0^\chi &= \lambda_{0,\rm r}^\chi + v \Delta^\chi + O(v^2), 
\label{lamda_expand}
\\
\hat{r}_0^\chi &= \hat{r}_{0,\rm r}^\chi + v \hat{\eta}^\chi + O(v^2), 
\label{r_expand}
\\
\hat{\ell}_0^\chi &= \hat{\ell}_{0,\rm r}^\chi + v \hat{\zeta}^\chi + O(v^2). 
\label{ell_expand}
\end{align} 
We note that $\lambda_{0,\rm r}^\chi$ is non-degenerate 
because the steady state is assumed to be uniquely determined for fixed $\bm{\alpha}$ in the RWA. 
We also note that $\hat{\ell}_{0,\rm r}^\chi$ and $\hat{r}_{0,\rm r}^\chi$ may have 
both the $O(v^0)$ and $O(v)$ terms, 
because $\mathcal{K}_{\rm r}^\chi$ has $O(v)$ terms as well as $O(v^0)$ ones.
Then, regarding $\mathcal{K}_{\rm r}^\chi$ as the unperturbed part and $v \mathcal{R}^\chi$ as the perturbation, 
we apply the formal perturbation theory for the non-degenerate case to obtain 
\begin{align}
\Delta^\chi &= {\rm Tr}_{\rm S} [\hat{\ell}_{0,{\rm r}}^{\chi\dag} \mathcal{R}^\chi \hat{r}_{0,\rm r}^\chi], 
\label{Delta}
\\
\hat{\eta}^\chi &= \sum_{m \neq 0} 
\frac{{\rm Tr}_{\rm S} [\hat{\ell}_{m,\rm r}^{\chi\dag} \mathcal{R}^\chi \hat{r}_{0,\rm r}^\chi]}
{\lambda_{0,\rm r}^\chi - \lambda_{m,\rm r}^\chi} \hat{r}_{m,\rm r}^\chi, 
\label{eta}
\\
\hat{\zeta}^\chi &= \sum_{m \neq 0} 
\left( \frac{{\rm Tr}_{\rm S} \bigl[ (\mathcal{R}^{\chi\dag} \hat{\ell}_{0,\rm r}^\chi)^\dag 
\hat{r}_{m,\rm r}^\chi \bigr]}{\lambda_{0,\rm r}^\chi - \lambda_{m,\rm r}^\chi} \right)^* 
\hat{\ell}_{m,\rm r}^\chi. 
\label{zeta}
\end{align}
The terms with $m=0$ vanish in Eqs.~(\ref{eta}) and (\ref{zeta}) 
because of the normalization condition ${\rm Tr}_{\rm S} [\hat{\ell}_0^{\chi\dag} \hat{r}_0^\chi]=1$.
Here, $\hat{\zeta}^\chi$ and $\hat{\eta}^\chi$ must be $O(v^0)$ 
for the ansatz and the formal perturbation theory to be consistent.
As we will show later, the denominators $\lambda_{0,\rm r}^\chi - \lambda_{m,\rm r}^\chi$ 
in Eqs.~(\ref{eta}) and (\ref{zeta}) are $O(v)$ for $m$'s 
where the corresponding eigenvectors $\hat{\ell}_{m,\rm r}^\chi$ and $\hat{r}_{m,\rm r}^\chi$ 
are in $\mathsf{P}$.
Therefore it is necessary to show that the corresponding numerators vanish.

We can show this as follows. 
We first note that $\mathcal{P} \mathcal{R}^\chi \mathcal{P}=0$ 
since $\mathcal{P} \mathcal{K}_{\rm r}^\chi\mathcal{P} = \mathcal{P} \mathcal{K}^\chi\mathcal{P}$ holds 
as is mentioned below Eq.~(\ref{coherencePartIs0_RWA}). 
This leads to ${\rm Tr}_{\rm S} [\hat{Y}_1^\dag \mathcal{R}^\chi \hat{Y}_2] = 0$ 
if $\hat{Y}_1, \hat{Y}_2 \in \mathsf{P}$.
Because $\hat{r}_{0,\rm r}^\chi \in \mathsf{P}$, 
we have ${\rm Tr}_{\rm S} [\hat{\ell}_{m,\rm r}^{\chi\dag} \mathcal{R}^\chi \hat{r}_{0,\rm r}^\chi] = 0$ 
if $\hat{\ell}_{m,\rm r}^\chi \in \mathsf{P}$. 
Similarly, we can show 
${\rm Tr}_{\rm S} [ (\mathcal{R}^{\chi\dag} \hat{\ell}_{0,\rm r}^\chi)^\dag \hat{r}_{m,\rm r}^\chi ] = 0$ 
if $\hat{r}_{m,\rm r}^\chi \in \mathsf{P}$. 
These indicate that the ansatz (\ref{lamda_expand})--(\ref{ell_expand}) 
and the formal perturbation theory do work. 

From this argument and Eq.~(\ref{Delta}), we have $\Delta^\chi=0$. 
Therefore from Eq.~(\ref{lamda_expand}) 
we obtain $\lambda_0^\chi = \lambda_{0,\rm r}^\chi + O(v^2)$. 
This implies that the analyses within and without the RWA give the equivalent result 
for the unit-time cumulant generating function for the entropy production 
in the steady state for fixed $\bm{\alpha}$. 

The above argument also implies that 
the nonzero contributions to $\hat{\zeta}^\chi$ and $\hat{\eta}^\chi$ 
come only from $m$'s which are in $\mathsf{Q}$.
Therefore we have 
\begin{align}
{\rm Tr}_{\rm S} \bigl[ \hat{\zeta}^{\chi\dag} \hat{r}_{0,\rm r}^\chi \bigr] 
= {\rm Tr}_{\rm S} \bigl[ \hat{\ell}_{0,\rm r}^{\chi\dag} \hat{\eta}^\chi \bigr] = 0. 
\label{ortho}
\end{align}
Substituting Eqs.~(\ref{r_expand}) and (\ref{ell_expand}) into Eq.~(\ref{vectorPotential}), 
we have 
\begin{align}
\bm{A}(\bm{\alpha}) = \bm{A}_{\rm r}(\bm{\alpha}) + v \bm{D} + O(v^2), 
\label{integrand0}
\end{align}
where 
\begin{align}
\bm{D} &:= - {\rm Tr}_{\rm S} \left[ \hat{\zeta}^{\prime\dag}(\bm{\alpha}) 
\frac{\partial \hat{\rho}_{\rm ss,r}(\bm{\alpha})}{\partial \bm{\alpha}} \right] 
- {\rm Tr}_{\rm S} \left[ \hat{\ell}^{\prime\dag}_{0,\rm r}(\bm{\alpha}) 
\frac{\partial \hat{\eta}(\bm{\alpha})}{\partial \bm{\alpha}} \right], 
\label{difference_NonRWA_RWA}
\end{align}
$\hat{\zeta}^\prime = \partial \hat{\zeta}^\chi/\partial (i\chi)|_{\chi=0}$, 
and $\hat{\eta} = \hat{\eta}^{\chi=0}$.
From Eq.~(\ref{ortho}), $\bm{D}$ vanishes. 
We thus derive Eq.~(\ref{integrand}), 
i.e., the results for the vector potential within and without the RWA are equivalent.

Finally, we show that $\lambda_{0,\rm r}^\chi - \lambda_{m,\rm r}^\chi = O(v)$ 
if and only if $\hat{\ell}_{m,\rm r}^\chi, \hat{r}_{m,\rm r}^\chi \in \mathsf{P}$.
Thanks to the decomposition of $\mathcal{K}^\chi_{\rm r}$ shown in Eq.~(\ref{K_r_chi}), 
we can expand its eigenvalue as 
$\lambda_{m,\rm r}^\chi = \lambda_{m,\rm r}^{(0)\chi} + v \lambda_{m,\rm r}^{(1)\chi} + O(v^2)$, 
and the eigenvectors as 
$\hat{r}_{m,\rm r}^\chi = \hat{r}_{m,\rm r}^{(0)\chi} + v \hat{r}_{m,\rm r}^{(1)\chi} + O(v^2)$. 
For $m$'s where $\hat{r}_{m,\rm r}^\chi \in \mathsf{P}$, 
we have $\mathcal{K}_0 \hat{r}_{m,\rm r}^{(0)\chi} = 0$, which implies $\lambda_{m,\rm r}^{(0)\chi} = 0$.
Therefore, for the eigenvalues whose eigenvectors are in $\mathsf{P}$, 
the difference $\lambda_{0,\rm r}^\chi - \lambda_{m,\rm r}^\chi$ is $O(v)$. 
On the other hand, for $m$'s where $\hat{r}_{m,\rm r}^\chi \in \mathsf{Q}$, 
$\mathcal{K}_0 \hat{r}_{m,\rm r}^{(0)\chi} \neq 0$ holds, 
so that $\lambda_{0,\rm r}^\chi - \lambda_{m,\rm r}^\chi$ is $O(v^0)$.

\subsection{Clausius Equality in Equilibrium State}\label{sec:equilibrium}

We next show that Eq.~(\ref{geometricalExcess}) reduces to the Clausius equality in the equilibrium setup.
In this setup, all the temperatures and chemical potentials of the reservoirs are equal:
$\beta_1=\beta_2=\cdots=:\beta$ and $\mu_1=\mu_2=\cdots=:\mu$, 
where $\beta$ and $\mu$ may be time-dependent. 
This situation is equivalent to the case where the system S is 
in contact with a single reservoir with the inverse temperature $\beta$ and chemical potential $\mu$.
Therefore we can omit the index $b$; 
for example, $\mathcal{K}_{\rm r}^\chi = \mathcal{K}_0 + v \mathcal{L}_{\rm r}^\chi$ 
and $\mathcal{K}_{\rm r}^\prime = v \mathcal{L}_{\rm r}^\prime$. 

In this case, we can show that the grand-canonical state is the steady solution of the QMME ($\chi=0$) 
within the RWA.
That is, $\mathcal{K}_{\rm r} \hat{\rho}_{\rm gc}(\beta,\beta\mu) = 0$, 
where $\hat{\rho}_{\rm gc}(\beta,\beta\mu) 
:= e^{-\beta\hat{H}_{\rm S} + \beta\mu\hat{N}_{\rm S}}/Z_{\rm gc}(\beta,\beta\mu)$ 
with $Z_{\rm gc}(\beta,\beta\mu) = {\rm Tr}_{\rm S} e^{-\beta\hat{H}_{\rm S} + \beta\mu\hat{N}_{\rm S}}$.

In order to have the explicit form of $\hat{\ell}^{\prime}_{0,\rm r}$ in the equilibrium setup, 
we differentiate the left eigenvalue equation 
$\mathcal{K}^{\chi\dag}_{\rm r} \hat{\ell}^\chi_{0,\rm r} = \lambda^{\chi *}_0 \hat{\ell}^\chi_{0,\rm r}$ 
with respect to $i\chi$ and set $\chi =0$:
\begin{align}
\mathcal{K}^{\prime\dag}_{\rm r} \hat{1} + \mathcal{K}^\dag_{\rm r} \hat{\ell}^{\prime}_{0,\rm r} = 0,
\label{equation_eq}
\end{align}
where we used $\lambda^{\chi *}_0 |_{\chi=0} = 0$, 
$\partial \lambda^{\chi *}_0 / \partial (i\chi)|_{\chi=0} = 0$,%
\footnote{
As is mentioned in Eq.~(\ref{entropyFlux}), 
$\partial \lambda^{\chi}_0 / \partial (i\chi)|_{\chi=0} = J_\sigma$ is the average entropy flow 
from the reservoirs into the system. 
This becomes zero in the equilibrium state because of the second law of thermodynamics.
}
and $\hat{\ell}_{0,\rm r}^{\chi=0} = \hat{1}$. 
To rewrite Eq.~(\ref{equation_eq}), 
we note that the following equation holds for any $\hat{Y} \in \mathsf{B}$:
\begin{align}
{\rm Tr}_{\rm S} \bigl[ (\mathcal{L}^{\prime\dag}_{\rm r} \hat{1})^\dag \hat{Y} \bigr] 
&= {\rm Tr}_{\rm S} \bigl[ (\mathcal{P} \mathcal{L}^{\prime\dag}_{\rm r} \mathcal{P} \hat{1})^\dag \hat{Y} \bigr] 
\notag\\
&= {\rm Tr}_{\rm S} \bigl[ \mathcal{P} \mathcal{L}^{\prime}_{\rm r} \mathcal{P} \hat{Y} \bigr] 
\notag\\
&= {\rm Tr}_{\rm S} \Bigl[ 
\beta (\hat{H}_{\rm S} -\mu \hat{N}_{\rm S}) \mathcal{P} \mathcal{L}_{\rm r} \mathcal{P} \hat{Y} 
- \mathcal{P} \mathcal{L}_{\rm r} \mathcal{P} \beta (\hat{H}_{\rm S} -\mu \hat{N}_{\rm S}) \hat{Y} \Bigr] 
\notag\\
&= {\rm Tr}_{\rm S} \Bigl[ 
\bigl\{ \mathcal{P} \mathcal{L}_{\rm r}^\dag \mathcal{P} 
\beta (\hat{H}_{\rm S} -\mu \hat{N}_{\rm S}) \bigr\}^\dag \hat{Y} 
- \bigl\{ \mathcal{P} \mathcal{L}_{\rm r}^\dag \mathcal{P} \hat{1} \bigr\}^\dag 
\beta (\hat{H}_{\rm S} -\mu \hat{N}_{\rm S}) \hat{Y} \Bigr] 
\notag\\
&= {\rm Tr}_{\rm S} \Bigl[ 
\bigl\{ \mathcal{L}_{\rm r}^\dag \beta (\hat{H}_{\rm S} -\mu \hat{N}_{\rm S}) \bigr\}^\dag \hat{Y} \Bigr],
\end{align}
where we used 
$\mathcal{P}\hat{1}=\hat{1}$ and $\mathcal{Q}\mathcal{L}^{\prime\dag}_{\rm r}\mathcal{P}=0$ in the first line, 
Eq.~(\ref{L_b_prime_L_b}) in the third line, $\mathcal{L}^\dag_{\rm r}\hat{1}=0$ in the fourth line, 
and $\mathcal{P}(\hat{H}_{\rm S} -\mu \hat{N}_{\rm S})=\hat{H}_{\rm S} -\mu \hat{N}_{\rm S}$ in the last line.
This equation implies 
\begin{align}
\mathcal{L}^{\prime\dag}_{\rm r} \hat{1} 
= \mathcal{L}_{\rm r}^\dag \beta (\hat{H}_{\rm S} -\mu \hat{N}_{\rm S}).
\label{L_prime_L}
\end{align} 
Therefore, we can rewrite Eq.~(\ref{equation_eq}) as 
\begin{align}
\mathcal{K}^\dag_{\rm r} \bigl\{ \hat{\ell}^{\prime}_{0,\rm r} 
+ \beta (\hat{H}_{\rm S} -\mu \hat{N}_{\rm S}) \bigr\} = 0.
\end{align}
Since the left eigenvector of $\mathcal{K}_{\rm r}$ corresponding to the zero eigenvalue 
is proportional to the identity operator $\hat{1}$, we have 
\begin{align}
\hat{\ell}^{\prime}_{0,\rm r} = - \beta (\hat{H}_{\rm S} -\mu \hat{N}_{\rm S}) + c\hat{1}, 
\label{ellPrime_eq}
\end{align}
where $c$ is an unimportant constant.

Substituting $\hat{\rho}_{\rm ss,r} = \hat{\rho}_{\rm gc}(\beta,\beta\mu)$ 
and Eq.~(\ref{ellPrime_eq}) into Eq.~(\ref{vectorPotentialRWA}), we obtain 
\begin{align}
\bm{A}_{\rm r}(\bm{\alpha}) &= \frac{\partial}{\partial \bm{\alpha}} S_{\rm vN}(\hat{\rho}_{\rm gc}) 
\label{integrand_equilibrium}
\end{align}
where 
$S_{\rm vN}(\hat{\rho}) := -{\rm Tr}_{\rm S} [ \hat{\rho} \ln \hat{\rho} ]$
is the von Neumann entropy of the state $\hat{\rho}$. 
In deriving Eq.~(\ref{integrand_equilibrium}), we used 
${\rm Tr}_{\rm S} \bigl[ \partial \hat{\rho}_{\rm ss,r}/\partial \bm{\alpha} \bigr] = 0$.

Finally, we note that the the grand-canonical state is the steady solution of QMME in the equilibrium setup 
also without the RWA: $\hat{\rho}_{\rm ss} = \hat{\rho}_{\rm gc}(\beta,\beta\mu)$.
Combining this fact with Eq.~(\ref{integrand}), we obtain the Clausius equality in the equilibrium setup 
(even without the RWA). 
That is, for the quasistatic operations in the equilibrium case, 
the change of the von Neumann entropy between the initial and final states 
is equal to the excess entropy production, 
which equals the total entropy production because the house-keeping part vanishes in the equilibrium states.

\subsection{Extended Clausius Equality in Weakly Nonequilibrium Regime}\label{sec:weak}

We now show that Eq.~(\ref{geometricalExcess}) reduces to 
the extended Clausius equality \cite{KNST1,KNST2,SaitoTasaki} in the weakly nonequilibrium setup.
In the later part of this subsection, we will show the following three equations.
\begin{align}
\bm{A}_{\rm r}(\bm{\alpha}) 
&= - {\rm Tr}_{\rm S} \left[ \ln \Bigl( \hat{\theta} \hat{\rho}_{\rm ss,r}(\bm{\alpha}) \hat{\theta}^{-1} \Bigr) 
\frac{\partial \hat{\rho}_{\rm ss,r}(\bm{\alpha})}{\partial \bm{\alpha}} \right] 
+ O(\epsilon^2), 
\label{integrand_weak}
\\
\frac{\partial}{\partial \bm{\alpha}} S_{\rm sym}\bigl( \hat{\rho}_{\rm ss,r}(\bm{\alpha}) \bigr) 
&= - {\rm Tr}_{\rm S} \left[ \ln \Bigl( \hat{\theta} \hat{\rho}_{\rm ss,r}(\bm{\alpha}) \hat{\theta}^{-1} \Bigr) 
\frac{\partial \hat{\rho}_{\rm ss,r}(\bm{\alpha})}{\partial \bm{\alpha}} \right] 
+ O(\epsilon^2), 
\label{symmetrizedSvN_RWA}
\\
S_{\rm sym}\bigl( \hat{\rho}_{\rm ss}(\bm{\alpha}) \bigr) 
&= S_{\rm sym}\bigl( \hat{\rho}_{\rm ss,r}(\bm{\alpha}) \bigr) + O(v^2), 
\label{symmetrizedSvN_RWA_nonRWA}
\end{align}
where $\epsilon$ is a measure of ``degree of nonequilibrium'' [see the below of Eq.~(\ref{epsilon})], and 
\begin{align}
S_{\rm sym}( \hat{\rho} ) 
:= - \frac{1}{2} {\rm Tr}_{\rm S} \Bigl[ \hat{\rho} 
\bigl( \ln \hat{\rho} + \ln \hat{\theta} \hat{\rho} \hat{\theta}^{-1} \bigr) \Bigr]
\label{symmetrizedSvN}
\end{align}
is the symmetrized von Neumann entropy \cite{SaitoTasaki}. 
$S_{\rm sym}$ is a quantum extension of the symmetrized Shannon entropy, 
which is first defined in Ref.~\cite{KNST1} in classical systems.

Equations (\ref{integrand_weak}) and (\ref{symmetrizedSvN_RWA}) lead to 
\begin{align}
\bm{A}_{\rm r}(\bm{\alpha}) 
&= \frac{\partial}{\partial \bm{\alpha}} S_{\rm sym}\bigl( \hat{\rho}_{\rm ss,r}(\bm{\alpha}) \bigr) 
+ O(\epsilon^2). 
\label{extendedClausiusRWA}
\end{align}
Combining this equation with Eqs.~(\ref{integrand}) and (\ref{symmetrizedSvN_RWA_nonRWA}), we have 
\begin{align}
\bm{A}(\bm{\alpha}) 
&= \frac{\partial}{\partial \bm{\alpha}} S_{\rm sym}\bigl( \hat{\rho}_{\rm ss}(\bm{\alpha}) \bigr) 
+ O(\epsilon^2) + O(v^2). 
\label{extendedClausius_nonRWA}
\end{align}
Equation (\ref{extendedClausius_nonRWA}) 
with the geometrical formula (\ref{geometricalExcess}) implies 
that the extended Clausius equality holds in the weakly nonequilibrium regime (even without the RWA): 
\begin{align}
\langle \sigma \rangle^{\rm ex} 
= S_{\rm sym}\bigl( \hat{\rho}_{\rm ss}(\bm{\alpha}^{\rm f}) \bigr)
- S_{\rm sym}\bigl( \hat{\rho}_{\rm ss}(\bm{\alpha}^{\rm i}) \bigr)
+ O(\epsilon^2 \delta) + O(v^2 \delta), 
\end{align}
where $\delta := \max_{\bm{\alpha}\in C} |\bm{\alpha} - \bm{\alpha}^{\rm i}| \big/ |\overline{\bm \alpha}|$, 
with $\overline{\bm \alpha}$ being a typical values of the control parameters.
In particular, if the initial state is an equilibrium state and 
the reservoir parameter change is included in the external operation, we have $\delta = O(\epsilon)$. 
Therefore the extended Clausius equality is valid up to $O(\epsilon^2)$ in this case.

Before going to the derivations, we make a remark that 
the symmetrized von Neumann entropy $S_{\rm sym}\bigl(\hat{\rho}_{\rm ss})$ 
of the steady state is equal to the von Neumann entropy 
$S_{\rm vN}\bigl(\hat{\rho}_{\rm ss}) = - {\rm Tr}_{\rm S} [\hat{\rho}_{\rm ss} \ln \hat{\rho}_{\rm ss}]$ 
if the spectrum of $\hat{H}_{\rm S}$ is non-degenerate.
We can show this as follows.
We first note that if $\hat{\rho}$ is time-reversal invariant, 
$\hat{\theta} \hat{\rho} \hat{\theta}^{-1} = \hat{\rho}$, 
then $S_{\rm sym}(\hat{\rho}) = S_{\rm vN}(\hat{\rho})$ holds.
If $\hat{H}_{\rm S}$ is non-degenerate, 
its eigenstates satisfy $\hat{\theta} | E_\nu \rangle = c_\nu | E_\nu \rangle$, 
where $c_\nu$ is a real constant with $c_\nu^2=1$, 
and the steady state $\hat{\rho}_{\rm ss,r}$ in the RWA is diagonal in the energy eigenstate basis. 
Therefore, $\hat{\theta} \hat{\rho}_{\rm ss,r} \hat{\theta}^{-1} = \hat{\rho}_{\rm ss,r}$ holds. 
This leads to $S_{\rm sym}(\hat{\rho}_{\rm ss,r}) = S_{\rm vN}(\hat{\rho}_{\rm ss,r})$. 
Furthermore, combining this result with Eqs.~(\ref{symmetrizedSvN_RWA_nonRWA}) and (\ref{SvN_RWA_nonRWA}), 
which will be shown later, 
we have $S_{\rm sym}(\hat{\rho}_{\rm ss}) = S_{\rm vN}(\hat{\rho}_{\rm ss}) + O(v^2)$.

\subsubsection*{Derivation of Eq.~(\ref{integrand_weak})}

In the weakly nonequilibrium regime, we can use a perturbative analysis 
with respect to the thermodynamic forces 
$\bigl\{ \epsilon_{1,b} := \beta_b - \overline{\beta} \bigr\}_b$ and 
$\bigl\{ \epsilon_{2,b} := \beta_b\mu_b - \overline{\beta\mu} \bigr\}_b$, 
where $\overline{\beta}$ and $\overline{\beta\mu}$ are reference values of the forces; 
we expand $\mathcal{K}^\chi_{\rm r}$, $\hat{\ell}^{\prime}_{0,\rm r}$, and $\hat{\rho}_{\rm ss,r}$ as 
\begin{align}
\mathcal{K}^\chi_{\rm r}(\bm{\alpha}) 
&= \overline{\mathcal{K}}^\chi_{\rm r} 
+ \sum_b \Bigl( \epsilon_{1,b} \overline{\partial_{1,b} \mathcal{K}}^\chi_{\rm r} 
+ \epsilon_{2,b} \overline{\partial_{2,b} \mathcal{K}}^\chi_{\rm r} \Bigr) + O(\epsilon^2), 
\label{expandKchi}
\\
\hat{\rho}_{\rm ss,r}(\bm{\alpha}) 
&= \overline{\rho}_{\rm ss,r} 
+ \sum_b \bigl( \epsilon_{1,b} \hat{\rho}_{1,b} + \epsilon_{2,b} \hat{\rho}_{2,b} \bigr) + O(\epsilon^2), 
\label{expandRho}
\\
\hat{\ell}^\prime_{0,\rm r}(\bm{\alpha}) 
&= \overline{\ell}^{\hspace{0.5mm}\prime}_{0,\rm r} 
+ \sum_b \bigl( \epsilon_{1,b} \hat{k}_{1,b} + \epsilon_{2,b} \hat{k}_{2,b} \bigr) + O(\epsilon^2), 
\label{expandEllPrime}
\end{align}
where 
$\overline{\mathcal{K}}^\chi_{\rm r} 
:= \mathcal{K}^\chi_{\rm r} \bigl( \bm{\alpha}_{\rm S}, \overline{\bm\alpha}_{\rm B} \bigr)$, 
$\overline{\rho}_{\rm ss,r} 
:= \hat{\rho}_{\rm ss,r} \bigl( \bm{\alpha}_{\rm S}, \overline{\bm\alpha}_{\rm B} \bigr)$, 
$\overline{\ell}^{\hspace{0.5mm}\prime}_{0,\rm r} 
:= \hat{\ell}^{\prime}_{0,\rm r} \bigl( \bm{\alpha}_{\rm S}, \overline{\bm\alpha}_{\rm B} \bigr)$, 
and 
\begin{align}
\overline{\partial_{i,b} \mathcal{K}}^\chi_{\rm r} 
&:= \frac{\partial \mathcal{K}^\chi_{\rm r}(\bm{\alpha})}{\partial \alpha_{i,b}}
\bigg|_{\bm{\alpha}_{\rm B} = \overline{\bm\alpha}_{\rm B}},
\label{epsilon}
\end{align}
with $i=1,2$, $\alpha_{1,b} = \beta_b$, and $\alpha_{2,b} = \beta_b\mu_b$.
Here, $\overline{\bm\alpha}_{\rm B}=\bigl\{ \overline{\beta}, \overline{\beta\mu} \bigr\}_b$ 
is the set of the reservoir parameters all of which are set to the reference values, 
$\epsilon= \max_b \epsilon_b$, and 
$\epsilon_b = \max \bigl\{ \epsilon_{1,b}/\overline{\beta}, \epsilon_{2,b}/\overline{\beta\mu} \bigr\}$. 
In the following we derive the relation between $\hat{\rho}_{\rm ss,r}$ and $\hat{\ell}^{\prime}_{0,\rm r}$ 
from the equations which determine $\overline{\rho}_{\rm ss,r}$, $\hat{\rho}_{i,b}$, 
$\overline{\ell}^{\hspace{0.5mm}\prime}_{0,\rm r}$, and $\hat{k}_{i,b}$.

We first investigate $\overline{\rho}_{\rm ss,r}$ and $\hat{\rho}_{i,b}$.
Substituting Eq.~(\ref{expandKchi}) with $\chi=0$ and Eq.~(\ref{expandRho}) 
into the right eigenvalue equation $\mathcal{K}_{\rm r}(\bm{\alpha}) \hat{\rho}_{\rm ss,r}(\bm{\alpha}) = 0$, 
we obtain 
\begin{align}
\overline{\mathcal{K}}_{\rm r} \overline{\rho}_{\rm ss,r} = 0
\label{equation_R_e0}
\end{align}
in $O(\epsilon^0)$, and 
\begin{align}
\overline{\mathcal{K}}_{\rm r} \hat{\rho}_{i,b} 
+ \overline{\partial_{i,b} \mathcal{K}}_{\rm r} \overline{\rho}_{\rm ss,r} 
= 0 
\label{equation_R_e1}
\end{align}
in $O(\epsilon_{i,b})$ with $i=1,2$. 
Because Eq.~(\ref{equation_R_e0}) is identical to the steady state equation for the equilibrium case, we have 
\begin{align}
\overline{\rho}_{\rm ss,r} = \hat{\rho}_{\rm gc} \bigl(\overline{\beta}, \overline{\beta\mu}).
\label{rho0}
\end{align}
Using this result, we rewrite the second term on the LHS of Eq.~(\ref{equation_R_e1}) as 
\begin{align}
\overline{\partial_{i,b} \mathcal{K}}_{\rm r} \overline{\rho}_{\rm ss,r} 
= v \overline{\mathcal{L}}_{b,\rm r} 
\left( \frac{\partial (\beta_b \hat{H}_{\rm S} - \beta_b\mu_b \hat{N}_{\rm S})}{\partial \alpha_{i,b}}
\bigg|_{\bm{\alpha}_{\rm B} = \overline{\bm\alpha}_{\rm B}} 
\hat{\rho}_{\rm gc} \bigl(\overline{\beta}, \overline{\beta\mu}) \right), 
\end{align}
where $\overline{\mathcal{L}}_{b,\rm r} 
:= \mathcal{L}_{b,\rm r} \bigl( \bm{\alpha}_{\rm S}, \overline{\bm\alpha}_{\rm B} \bigr)$.
We can derive this equation from the $\alpha_{i,b}$-derivative of the eigenvalue equation 
$\mathcal{L}_{b,\rm r} \hat{\rho}_{\rm gc} \bigl(\beta_b, \beta_b \mu_b) = 0$. 
Therefore we can rewrite Eq.~(\ref{equation_R_e1}) as 
\begin{align}
\overline{\mathcal{K}}_{\rm r} \hat{\rho}_{1,b} 
+ v \overline{\mathcal{L}}_{b,\rm r} 
\Bigl( \hat{H}_{\rm S} \hat{\rho}_{\rm gc} \bigl(\overline{\beta}, \overline{\beta\mu}) \Bigr) &= 0, 
\label{rho_1b}
\\
\overline{\mathcal{K}}_{\rm r} \hat{\rho}_{2,b} 
- v \overline{\mathcal{L}}_{b,\rm r} 
\Bigl( \hat{N}_{\rm S} \hat{\rho}_{\rm gc} \bigl(\overline{\beta}, \overline{\beta\mu}) \Bigr) &= 0. 
\label{rho_2b}
\end{align}
To proceed further, we note that the following relation holds for any $\hat{Y} \in \mathsf{B}$: 
\begin{align}
\overline{\mathcal{L}}_{b,\rm r} 
\Bigl( \hat{Y} \hat{\rho}_{\rm gc} \bigl(\overline{\beta}, \overline{\beta\mu}) \Bigr) 
&= \Bigl( \overline{\mathcal{L}}_{b,\rm r}^\dag \hat{Y} \Bigr) 
\hat{\rho}_{\rm gc} \bigl(\overline{\beta}, \overline{\beta\mu}). 
\label{detailedBalance1}
\end{align}
Equation~(\ref{detailedBalance1}) is the detailed balance condition for the QMME \cite{SpohnLebowitz}.
We can derive this equation with the help of the KMS condition (\ref{KMS}).
By using Eq.~(\ref{detailedBalance1}), we rewrite Eqs.~(\ref{rho_1b}) and (\ref{rho_2b}) as 
\begin{align}
\overline{\mathcal{K}}_{\rm r} \hat{\rho}_{1,b} 
+ v \Bigl( \overline{\mathcal{L}}_{b,\rm r}^\dag \hat{H}_{\rm S} \Bigr) 
\hat{\rho}_{\rm gc} \bigl(\overline{\beta}, \overline{\beta\mu}) = 0, 
\\
\overline{\mathcal{K}}_{\rm r} \hat{\rho}_{2,b} 
- v \Bigl( \overline{\mathcal{L}}_{b,\rm r}^\dag \hat{N}_{\rm S} \Bigr) 
\hat{\rho}_{\rm gc} \bigl(\overline{\beta}, \overline{\beta\mu}) = 0. 
\end{align}
Multiplying $\hat{\rho}_{\rm gc}^{-1} \bigl(\overline{\beta}, \overline{\beta\mu})$ 
from the right and taking the time reversal, we obtain 
\begin{align}
\hat{\theta} \Bigl( \overline{\mathcal{K}}_{\rm r} \hat{\rho}_{1,b} \Bigr) 
\hat{\theta}^{-1} \hat{\rho}_{\rm gc}^{-1} \bigl(\overline{\beta}, \overline{\beta\mu}) 
+ v \overline{\mathcal{L}}_{b,\rm r}^\dag \hat{H}_{\rm S} 
= 0, 
\label{rho_1b_2}
\\
\hat{\theta} \Bigl( \overline{\mathcal{K}}_{\rm r} \hat{\rho}_{2,b} \Bigr) 
\hat{\theta}^{-1} \hat{\rho}_{\rm gc}^{-1} \bigl(\overline{\beta}, \overline{\beta\mu}) 
- v \overline{\mathcal{L}}_{b,\rm r}^\dag \hat{N}_{\rm S} 
= 0,
\label{rho_2b_2} 
\end{align}
where we used $\widetilde{\mathcal{L}_{b,\rm r}^\dag} = \mathcal{L}_{b,\rm r}^\dag$, 
$\tilde{H}_{\rm S} = \hat{H}_{\rm S}$, and $\tilde{N}_{\rm S} = \hat{N}_{\rm S}$.
Furthermore, we can rewrite the first terms on the LHS as
\begin{align}
\hat{\theta} \Bigl( \overline{\mathcal{K}}_{\rm r} 
\hat{\rho}_{i,b} \Bigr) \hat{\theta}^{-1} \hat{\rho}_{\rm gc}^{-1} \bigl(\overline{\beta}, \overline{\beta\mu}) 
= \overline{\mathcal{K}}_{\rm r}^\dag 
\Bigl( \hat{\theta} \hat{\rho}_{i,b} \hat{\theta}^{-1} 
\hat{\rho}_{\rm gc}^{-1} \bigl(\overline{\beta}, \overline{\beta\mu}) \Bigr).
\end{align}
We can derive this equation from Eq.~(\ref{detailedBalance1}).
Therefore, we can rewrite Eqs.~(\ref{rho_1b_2}) and (\ref{rho_2b_2}) as 
\begin{align}
\overline{\mathcal{K}}_{\rm r}^\dag 
\Bigl( \hat{\theta} \hat{\rho}_{1,b} \hat{\theta}^{-1} 
\hat{\rho}_{\rm gc}^{-1} \bigl(\overline{\beta}, \overline{\beta\mu}) \Bigr) 
+ v \overline{\mathcal{L}}_{b,\rm r}^\dag \hat{H}_{\rm S} 
= 0, 
\label{rho_1b_3}
\\
\overline{\mathcal{K}}_{\rm r}^\dag 
\Bigl( \hat{\theta} \hat{\rho}_{2,b} \hat{\theta}^{-1} 
\hat{\rho}_{\rm gc}^{-1} \bigl(\overline{\beta}, \overline{\beta\mu}) \Bigr) 
- v \overline{\mathcal{L}}_{b,\rm r}^\dag \hat{N}_{\rm S} 
= 0.
\label{rho_2b_3} 
\end{align}

We next investigate $\overline{\ell}^{\hspace{0.5mm}\prime}_{0,\rm r}$ and $\hat{k}_{i,b}$.
We differentiate the left eigenvalue equation of $\mathcal{K}^\chi_{\rm r}$ with respect to $i\chi$
and set $\chi =0$:
\begin{align}
\mathcal{K}^{\prime\dag}_{\rm r}(\bm{\alpha}) \hat{1} 
+ \mathcal{K}^\dag_{\rm r}(\bm{\alpha}) \hat{\ell}^{\hspace{0.5mm}\prime}_{0,\rm r}(\bm{\alpha}) 
= -J_\sigma(\bm{\alpha}) \hat{1},
\label{equationForEllPrime}
\end{align}
where we used $\lambda^{\chi *}_0 |_{\chi=0} = 0$, 
$\partial \lambda^{\chi *}_0 / \partial (i\chi)|_{\chi=0} = -J_\sigma$, 
and $\hat{\ell}_0^{\chi=0} = \hat{1}$. 
In the NESSs close to equilibrium, we can write the average entropy flow $J_\sigma$ in a quadratic form:
$J_\sigma = \sum_{i,j=1,2} \sum_{b,b'} L_{i,b;j,b'} \epsilon_{i,b} \epsilon_{j,b'}$ 
with a phenomenological coefficient $L_{i,b;j,b'}$. 
This implies that the RHS of Eq.~(\ref{equationForEllPrime}) is $O(\epsilon^2)$.
By substituting Eqs.~(\ref{expandKchi}) and (\ref{expandEllPrime}) into Eq.~(\ref{equationForEllPrime}), 
we obtain 
\begin{align}
\overline{\mathcal{K}}_{\rm r}^{\hspace{0.6mm}\prime\dag} \hat{1} 
+ \overline{\mathcal{K}}_{\rm r}^\dag \overline{\ell}^{\hspace{0.5mm}\prime}_{0,\rm r} 
=0
\label{equation_e0}
\end{align}
in $O(\epsilon^0)$, and 
\begin{align}
\overline{\partial_{i,b} \mathcal{K}}^{\hspace{.6mm}\prime\dag}_{\rm r} \hat{1} 
+ \overline{\mathcal{K}}_{\rm r}^\dag \hat{k}_{i,b} 
+ \overline{\partial_{i,b} \mathcal{K}}^\dag_{\rm r} \overline{\ell}^{\hspace{0.5mm}\prime}_{0,\rm r} 
= 0 
\label{equation_e1}
\end{align}
in $O(\epsilon_{i,b})$ with $i=1,2$.
Here, 
$\overline{\mathcal{K}}_{\rm r}^{\hspace{0.6mm}\prime} 
:= \mathcal{K}_{\rm r}^\prime \bigl( \bm{\alpha}_{\rm S}, \overline{\bm\alpha}_{\rm B} \bigr)$, 
and 
$\overline{\partial_{i,b} \mathcal{K}}^{\hspace{0.6mm}\prime}_{\rm r} 
:= \bigl( \partial \mathcal{K}^\prime_{\rm r}(\bm{\alpha}) \big/ \partial \alpha_{i,b} \bigr) 
\big|_{\bm{\alpha}_{\rm B} = \overline{\bm\alpha}_{\rm B}}$. 
Equation~(\ref{equation_e0}) is identical to Eq.~(\ref{equation_eq}), so that
\begin{align}
\overline{\ell}^{\hspace{0.5mm}\prime}_{0,\rm r} 
=  - \overline{\beta} \hat{H}_{\rm S} + \overline{\beta\mu} \hat{N}_{\rm S} + c\hat{1}. 
\label{ellPrime0_e}
\end{align}
We rewrite the first term on the LHS of Eq.~(\ref{equation_e1}) as 
\begin{align}
\overline{\partial_{i,b} \mathcal{K}}^{\hspace{0.6mm}\prime\dag}_{\rm r} \hat{1} 
&= v \frac{\partial \bigl( \mathcal{L}^{\prime\dag}_{b,\rm r}(\bm{\alpha})\hat{1}\bigr)}
{\partial \alpha_{i,b}}\bigg|_{\bm{\alpha}_{\rm B} = \overline{\bm\alpha}_{\rm B}} 
\notag\\
&= v \frac{\partial \bigl[ \mathcal{L}^{\dag}_{b,\rm r}(\bm{\alpha}) 
(\beta_b \hat{H}_{\rm S} - \beta_b\mu_b \hat{N}_{\rm S}) \bigr]}{\partial \alpha_{i,b}}
\bigg|_{\bm{\alpha}_{\rm B} = \overline{\bm\alpha}_{\rm B}} 
\notag\\
&= v \overline{\partial_{i,b} \mathcal{L}}^{\dag}_{b,\rm r}
(\overline{\beta} \hat{H}_{\rm S} - \overline{\beta\mu} \hat{N}_{\rm S})
+ v \overline{\mathcal{L}}^\dag_{b,\rm r} 
\frac{\partial (\beta_b \hat{H}_{\rm S} - \beta_b\mu_b \hat{N}_{\rm S})}{\partial \alpha_{i,b}}
\bigg|_{\bm{\alpha}_{\rm B} = \overline{\bm\alpha}_{\rm B}}, 
\label{firstTerm}
\end{align}
where $\overline{\partial_{i,b} \mathcal{L}}_{b,\rm r} 
:= \bigl( \partial \mathcal{L}_{b,\rm r}(\bm{\alpha}) \big/ \partial \alpha_{i,b} \bigr)
\big|_{\bm{\alpha}_{\rm B} = \overline{\bm\alpha}_{\rm B}}$.
Here, we used Eq.~(\ref{L_prime_L}) in the second line. 
By using Eq.~(\ref{ellPrime0_e}), we also rewrite the third term on the LHS of Eq.~(\ref{equation_e1}) as 
\begin{align}
\overline{\partial_{i,b} \mathcal{K}}^\dag_{\rm r} 
\bigl( - \overline{\beta} \hat{H}_{\rm S} + \overline{\beta\mu} \hat{N}_{\rm S} + c\hat{1} \bigr) 
= - v \overline{\partial_{i,b} \mathcal{L}}^\dag_{b,\rm r} 
\bigl(\overline{\beta} \hat{H}_{\rm S} - \overline{\beta\mu} \hat{N}_{\rm S} \bigr), 
\label{thirdTerm}
\end{align}
where we used the fact that 
$\overline{\partial_{i,b} \mathcal{K}}^\dag_{\rm r} \hat{1} = 0$ holds 
(we can derive this by differentiating the left eigenvalue equation $\mathcal{K}_{\rm r}^\dag \hat{1} = 0$ 
with respect to $\alpha_{i,b}$).
Equation~(\ref{thirdTerm}) cancels out the first term of Eq.~(\ref{firstTerm}).
Thus Eq.~(\ref{equation_e1}) yields 
\begin{align}
\overline{\mathcal{K}}^\dag_{\rm r} \hat{k}_{1,b} 
+ v \overline{\mathcal{L}}^\dag_{b,\rm r} \hat{H}_{\rm S} = 0,
\label{k_1b}
\\
\overline{\mathcal{K}}^\dag_{\rm r} \hat{k}_{2,b} 
- v \overline{\mathcal{L}}^\dag_{b,\rm r} \hat{N}_{\rm S} = 0.
\label{k_2b}
\end{align}

Then, combining Eq.~(\ref{rho_1b_3}) with Eq.~(\ref{k_1b}) and Eq.~(\ref{rho_2b_3}) with Eq.~(\ref{k_2b}), 
we obtain 
\begin{align}
\overline{\mathcal{K}}^\dag_{\rm r} 
\Bigl( \hat{k}_{i,b} - \hat{\theta} \hat{\rho}_{i,b} \hat{\theta}^{-1} 
\hat{\rho}_{\rm gc}^{-1} (\overline{\beta}, \overline{\beta\mu}) \Bigr) = 0, 
\end{align}
for $i=1,2$. 
This results in 
\begin{align}
\hat{k}_{i,b} 
= \hat{\theta} \hat{\rho}_{i,b} \hat{\theta}^{-1} 
\hat{\rho}_{\rm gc}^{-1} (\overline{\beta}, \overline{\beta\mu}) 
+ c_{i,b} \hat{1},
\label{k_ib}
\end{align}
where $c_{i,b}$ is a constant.
Combining the results of Eqs.~(\ref{expandRho}), (\ref{expandEllPrime}), 
(\ref{rho0}), (\ref{ellPrime0_e}), and (\ref{k_ib}), we obtain 
\begin{align}
\hat{\ell}^{\prime}_{0,\rm r}(\bm{\alpha}) 
&= \ln \hat{\rho}_{\rm gc} (\overline{\beta}, \overline{\beta\mu}) 
+ \sum_b \bigl( \epsilon_{1,b} \hat{\theta} \hat{\rho}_{1,b} \hat{\theta}^{-1}  
+ \epsilon_{2,b} \hat{\theta} \hat{\rho}_{2,b} \hat{\theta}^{-1}  \bigr)
\hat{\rho}_{\rm gc}^{-1} (\overline{\beta}, \overline{\beta\mu}) 
+ c \hat{1} + O(\epsilon^2)
\notag\\
&= \ln \hat{\theta} \hat{\rho}_{\rm ss,r}(\bm{\alpha}) \hat{\theta}^{-1} + c \hat{1} + O(\epsilon^2), 
\end{align}
where $c$ is a constant.
Finally, substituting this result into Eq.~(\ref{vectorPotentialRWA}), 
we obtain Eq.~(\ref{integrand_weak}): 
\begin{align*}
\bm{A}_{\rm r}(\bm{\alpha})
= - {\rm Tr}_{\rm S} \left[ 
\Bigl( \ln \hat{\theta} \hat{\rho}_{\rm ss,r}(\bm{\alpha}) \hat{\theta}^{-1} \Bigr) 
\frac{\partial \hat{\rho}_{\rm ss,r}(\bm{\alpha})}{\partial \bm{\alpha}} \right] 
+ O(\epsilon^2), 
\end{align*}
where we used 
${\rm Tr}_{\rm S} \bigl[ \partial \hat{\rho}_{\rm ss,r}/\partial \bm{\alpha} \bigr] = 0$.

\subsubsection*{Derivation of Eq.~(\ref{symmetrizedSvN_RWA})}

The difference between the RHS and LHS of Eq.~(\ref{symmetrizedSvN_RWA}) is written as
\begin{align}
&- \frac{\partial}{\partial \bm{\alpha}} S_{\rm sym}\bigl( \hat{\rho}_{\rm ss,r}(\bm{\alpha}) \bigr) 
- {\rm Tr}_{\rm S} \left[ 
\frac{\partial \hat{\rho}_{\rm ss,r}(\bm{\alpha})}{\partial \bm{\alpha}} 
\ln \tilde{\rho}_{\rm ss,r}(\bm{\alpha}) \right]
\notag\\
&= \frac{1}{2}{\rm Tr}_{\rm S} \left[ \frac{\partial \hat{\rho}_{\rm ss,r}}{\partial \bm{\alpha}} 
\Bigl( \ln \hat{\rho}_{\rm ss,r} - \ln \tilde{\rho}_{\rm ss,r} \Bigr) \right] 
+ \frac{1}{2} {\rm Tr}_{\rm S} \left[ 
\frac{\partial \hat{\rho}_{\rm ss,r}}{\partial \bm{\alpha}} \right] 
+ \frac{1}{2} {\rm Tr}_{\rm S} \left[ 
\frac{\partial \tilde{\rho}_{\rm ss,r}}{\partial \bm{\alpha}} 
\hat{\rho}_{\rm ss,r} \tilde{\rho}_{\rm ss,r}^{-1} \right]. 
\label{del_S_sym}
\end{align}
The second term on the RHS vanishes because of the normalization condition. 
From Eq.~(\ref{expandRho}) and the time-reversal invariance of 
$\overline{\rho}_{\rm ss,r} = \hat{\rho}_{\rm gc} (\overline{\beta}, \overline{\beta\mu})$, 
we can show $\hat{\rho}_{\rm ss,r} - \tilde{\rho}_{\rm ss,r} = \epsilon \hat{\psi} + O(\epsilon^2)$.
Therefore, we have 
\begin{align}
\ln \hat{\rho}_{\rm ss,r} - \ln \tilde{\rho}_{\rm ss,r} 
&= \ln \bigl( \tilde{\rho}_{\rm ss,r} + \epsilon \hat{\psi} \bigr) 
- \ln \tilde{\rho}_{\rm ss,r} + O(\epsilon^2) 
\notag\\
&= \epsilon \hat{\psi} \tilde{\rho}_{\rm ss,r}^{-1} + O(\epsilon^2). 
\end{align}
Using this result, we can evaluate the first term on the RHS of (\ref{del_S_sym}) as 
\begin{align}
\frac{1}{2}{\rm Tr}_{\rm S} \left[ \frac{\partial \hat{\rho}_{\rm ss,r}}{\partial \bm{\alpha}} 
\Bigl( \ln \hat{\rho}_{\rm ss,r} - \ln \tilde{\rho}_{\rm ss,r} \Bigr) \right] 
&= \frac{\epsilon}{2}{\rm Tr}_{\rm S} \left[ \frac{\partial \hat{\rho}_{\rm ss,r}}{\partial \bm{\alpha}} 
\hat{\psi} \tilde{\rho}_{\rm ss,r}^{-1} \right] + O(\epsilon^2) 
\notag\\
&= \frac{\epsilon}{2}{\rm Tr}_{\rm S} \left[ 
\frac{\partial (\hat{\theta}\hat{\rho}_{\rm ss,r}\hat{\theta}^{-1})}{\partial \bm{\alpha}} 
\hat{\theta}\hat{\psi}\hat{\theta}^{-1} \hat{\rho}_{\rm ss,r}^{-1} \right] + O(\epsilon^2) 
\notag\\
&= - \frac{\epsilon}{2}{\rm Tr}_{\rm S} \left[ \frac{\partial \tilde{\rho}_{\rm ss,r}}{\partial \bm{\alpha}} 
\hat{\psi} \hat{\rho}_{\rm ss,r}^{-1} \right] + O(\epsilon^2), 
\end{align}
where we used that ${\rm Tr}_{\rm S}\hat{Y} = {\rm Tr}_{\rm S}\hat{\theta}\hat{Y}\hat{\theta}^{-1}$ 
if ${\rm Tr}_{\rm S}\hat{Y}$ is real in the second line, 
and $\hat{\theta}\hat{\psi}\hat{\theta}^{-1} = \tilde{\psi} = - \hat{\psi}$ in the third line.
We can evaluate the third term on the RHS of (\ref{del_S_sym}) as 
\begin{align}
\frac{1}{2}{\rm Tr}_{\rm S} \left[ \frac{\partial \tilde{\rho}_{\rm ss,r}}{\partial \bm{\alpha}} 
\hat{\rho}_{\rm ss,r} \tilde{\rho}_{\rm ss,r}^{-1} \right] 
&= \frac{1}{2}{\rm Tr}_{\rm S} \left[ \frac{\partial \tilde{\rho}_{\rm ss,r}}{\partial \bm{\alpha}} 
\bigl( \tilde{\rho}_{\rm ss,r} + \epsilon \hat{\psi} \bigr) \tilde{\rho}_{\rm ss,r}^{-1} \right] 
+ O(\epsilon^2)
\notag\\
&= \frac{\epsilon}{2}{\rm Tr}_{\rm S} \left[ \frac{\partial \tilde{\rho}_{\rm ss,r}}{\partial \bm{\alpha}} 
\hat{\psi} \hat{\rho}_{\rm ss,r}^{-1} \right] 
+ O(\epsilon^2), 
\end{align}
where in the second line we used 
${\rm Tr}_{\rm S} \bigl[ \partial \tilde{\rho}_{\rm ss,r} \big/ \partial \bm{\alpha} \bigr] = 0$
and $\epsilon \tilde{\rho}_{\rm ss,r}^{-1} = \epsilon \hat{\rho}_{\rm ss,r}^{-1} + O(\epsilon^2)$.

Substituting these results into the first and third terms on the RHS of (\ref{del_S_sym}), 
we obtain 
\begin{align}
\frac{\partial}{\partial \bm{\alpha}} S_{\rm sym}\bigl( \hat{\rho}_{\rm ss,r}(\bm{\alpha}) \bigr) 
= - {\rm Tr}_{\rm S} \left[ \frac{\partial \hat{\rho}_{\rm ss,r}(\bm{\alpha})}{\partial \bm{\alpha}} 
\ln \tilde{\rho}_{\rm ss,r}(\bm{\alpha}) \right] 
+ O(\epsilon^2). 
\end{align}

\subsubsection*{Derivation of Eq.~(\ref{symmetrizedSvN_RWA_nonRWA})}

We note that the symmetrized von Neumann entropy is written as 
\begin{align}
S_{\rm sym}(\hat{\rho}) = \frac{1}{2} \bigl\{ S_{\rm vN}(\hat{\rho}) + \tilde{S}(\hat{\rho}) \bigr\}, 
\end{align}
where $\tilde{S}(\hat{\rho}) := - {\rm Tr}_{\rm S} \bigl[ \hat{\rho} \ln \tilde{\rho} \bigr]$.
In the following, we derive 
\begin{align}
S_{\rm vN}(\hat{\rho}_{\rm ss}) &= S_{\rm vN}(\hat{\rho}_{\rm ss,r}) + O(v^2),
\label{SvN_RWA_nonRWA}
\\ 
\tilde{S}(\hat{\rho}_{\rm ss}) &= \tilde{S}(\hat{\rho}_{\rm ss,r}) + O(v^2). 
\label{tildeS_RWA_nonRWA}
\end{align}
These lead to Eq.~(\ref{symmetrizedSvN_RWA_nonRWA}).

First, we consider $S_{\rm vN}$. 
We represent $\hat{\rho}_{\rm ss,r}$ by a matrix in the basis of the eigenstates of $\hat{H}_{\rm S}$.
Then, because $\hat{\rho}_{\rm ss,r}$ is in $\mathsf{P}$, 
$\hat{\rho}_{\rm ss,r}$ is represented by a block diagonal matrix. 
Each block is in the degenerate subspace that has a single energy eigenvalue; i.e., 
the non-vanishing matrix elements are $\{ \langle E_\nu, n | \hat{\rho}_{\rm ss,r} | E_\nu, n' \rangle \}_\nu$.
Therefore, if we take the appropriate linear combination $| E_\nu, l \rangle$ of $\{ | E_\nu, n \rangle \}_n$, 
which is also an energy eigenstate, in each degenerate subspace, 
we can diagonalize $\hat{\rho}_{\rm ss,r}$.
Using this diagonalizing basis, we have 
\begin{align}
S_{\rm vN}(\hat{\rho}_{\rm ss,r}) 
= - \sum_{\nu,l} \bigl[ \hat{\rho}_{\rm ss,r} \bigr]_{\nu,l} 
\ln \bigl[ \hat{\rho}_{\rm ss,r} \bigr]_{\nu,l}, 
\label{SvN_RWA}
\end{align}
where $\bigl[ \hat{\rho}_{\rm ss,r} \bigr]_{\nu,l} 
:= \langle E_\nu, l | \hat{\rho}_{\rm ss,r} | E_\nu, l \rangle$.
We assume that there is no degeneracy in the eigenvalues of $\hat{\rho}_{\rm ss,r}$; 
i.e., $\bigl[ \hat{\rho}_{\rm ss,r} \bigr]_{\nu,l} \neq \bigl[ \hat{\rho}_{\rm ss,r} \bigr]_{\nu',l'}$ 
if $\nu\neq\nu'$ or $l \neq l'$. 

As we showed in Sec~\ref{sec:main}, we can write the steady state without the RWA as 
\begin{align}
\hat{\rho}_{\rm ss} = \hat{\rho}_{\rm ss,r} + v \hat{\eta} + O(v^2). 
\label{rho_ss}
\end{align}
We evaluate the eigenvalue $e_{\nu,l}$ of $\hat{\rho}_{\rm ss}$ 
by regarding $\hat{\rho}_{\rm ss,r}$ as the unperturbed part and $v \hat{\eta}$ as the perturbation: 
\begin{align}
e_{\nu,l} = \bigl[ \hat{\rho}_{\rm ss,r} \bigr]_{\nu,l; \nu,l} + v \Delta_{\nu,l} + O(v^2).
\end{align}
Because we assume the non-degeneracy in the eigenvalues of $\hat{\rho}_{\rm ss,r}$,
we can use the perturbation theory for the non-degenerate case to obtain 
\begin{align}
\Delta_{\nu,l} = \langle E_\nu,l | \hat{\eta} | E_\nu,l \rangle. 
\end{align}
Because $\hat{\eta} \in \mathsf{Q}$, as we showed in the previous subsection, $\Delta_{\nu,l}$ vanishes.
Therefore we obtain 
\begin{align}
S_{\rm vN}(\hat{\rho}_{\rm ss}) &= - \sum_{\nu,l} e_{\nu,l} \ln e_{\nu,l} 
\notag\\
&= - \sum_{\nu,l} \bigl[ \hat{\rho}_{\rm ss,r} \bigr]_{\nu,l} 
\ln \bigl[ \hat{\rho}_{\rm ss,r} \bigr]_{\nu,l} + O(v^2).   
\end{align}
Comparing this with Eq.~(\ref{SvN_RWA}), we derive Eq.~(\ref{SvN_RWA_nonRWA}).

Next, we consider $\tilde{S}$.
We note that $\tilde{\rho}_{\rm ss,r}$ is the steady solution of $\tilde{\mathcal{K}}_{\rm r}$; 
i.e., $\tilde{\mathcal{K}}_{\rm r} \tilde{\rho}_{\rm ss,r}=0$. 
Similarly to the case of $\hat{\rho}_{\rm ss,r}$ we can show $\tilde{\rho}_{\rm ss,r} \in \mathsf{P}$.
Therefore by the same argument as the above, we can diagonalize $\tilde{\rho}_{\rm ss,r}$ 
by taking the appropriate energy eigenstates $|E_\nu, \tilde{l} \rangle$. 
We note that $\{ |E_\nu, \tilde{l} \rangle \}_{\tilde{l}}$ is different from 
$\{ |E_\nu, l \rangle \}_l$ in the above, 
and that $\hat{\rho}_{\rm ss,r}$ is not diagonalized 
in the basis of $\{ |E_\nu, \tilde{l} \rangle \}_{\tilde{l}}$.
Using this basis, we have 
\begin{align}
\tilde{S}(\hat{\rho}_{\rm ss,r}) 
= - \sum_{\nu,\tilde{l}} \bigl[ \hat{\rho}_{\rm ss,r} \bigr]_{\nu,\tilde{l}} 
\ln \bigl[ \tilde{\rho}_{\rm ss,r} \bigr]_{\nu,\tilde{l}}, 
\label{tildeS_RWA}
\end{align}
where $\bigl[ \tilde{\rho}_{\rm ss,r} \bigr]_{\nu,\tilde{l}} 
= \langle E_\nu,\tilde{l}| \tilde{\rho}_{\rm ss,r} | E_\nu,\tilde{l}\rangle$ 
and $\bigl[ \hat{\rho}_{\rm ss,r} \bigr]_{\nu,\tilde{l}} 
= \langle E_\nu,\tilde{l}|\hat{\rho}_{\rm ss,r} | E_\nu,\tilde{l}\rangle$.

The time reversal $\tilde{\rho}_{\rm ss}$ of the steady state $\hat{\rho}_{\rm ss}$ without the RWA 
is the steady solution of $\tilde{\mathcal{K}}$. 
Taking the time reversal of Eq.~(\ref{rho_ss}), we have 
\begin{align}
\tilde{\rho}_{\rm ss} = \tilde{\rho}_{\rm ss,r} + v \tilde{\eta} + O(v^2). 
\end{align}
By an argument similar to that in the previous subsection, we can show $\tilde{\eta} \in \mathsf{Q}$. 
Therefore, as in the above, we can evaluate the eigenvalue $\tilde{e}_{\nu,\tilde{l}}$ 
of $\tilde{\rho}_{\rm ss}$ as 
\begin{align}
\tilde{e}_{\nu,\tilde{l}} = \bigl[ \tilde{\rho}_{\rm ss,r} \bigr]_{\nu,\tilde{l}} + O(v^2).
\label{e_nu_tilde_l}
\end{align}
Therefore we obtain 
\begin{align}
\tilde{S}(\hat{\rho}_{\rm ss}) 
&= - \sum_{\nu,\tilde{l}} \bigl[ \hat{\rho}_{\rm ss,r} \bigr]_{\nu,\tilde{l}} \ln \tilde{e}_{\nu,\tilde{l}} 
\notag\\
&= - \sum_{\nu,\tilde{l}} \bigl[ \hat{\rho}_{\rm ss,r} \bigr]_{\nu,\tilde{l}} 
\ln \bigl[ \tilde{\rho}_{\rm ss,r} \bigr]_{\nu,\tilde{l}} + O(v^2).   
\end{align}
Comparing this with Eq.~(\ref{tildeS_RWA}), we derive Eq.~(\ref{tildeS_RWA_nonRWA}).

\section{Example: Spinless Electron System in Quantum Dots}\label{sec:examples}

In this section, we investigate the excess entropy production for quasistatic operations 
in a simple electron model to demonstrate the general results in the previous section. 
We consider a spinless electron system in $N$ quantum dots connected to $N_{\rm B}$ electron reservoirs. 
An example of the system with $N=4$ and $N_{\rm B}=2$ is illustrated in Fig.~\ref{fig:example}. 
We assume that each dot has a single level.
The Hamiltonian of the total system is given in the form of Eq.~(\ref{Htot}), where  
\begin{align}
\hat{H}_{\rm S} &= \sum_{i=1}^N \varepsilon_i \hat{d}_i^\dag \hat{d}_i 
+ \sum_{\langle ii'\rangle} t_{ii'} (\hat{d}_i^\dag \hat{d}_{i'} + {\rm h.c.}) 
+ U \sum_{\langle ii'\rangle} \hat{d}_i^\dag \hat{d}_i \hat{d}_{i'}^\dag \hat{d}_{i'},  
\label{HamiltonianDot}
\\
\hat{H}_b &= \sum_k \hbar\Omega_{bk} \hat{c}_{bk}^\dag \hat{c}_{bk},
\\
\hat{H}_{{\rm S}b} &= \sum_k \sum_{i=1}^N \xi_{ibk} (\hat{d}_i^\dag \hat{c}_{bk} + {\rm h.c.}). 
\label{coupleHamiltonianDot}
\end{align}
Here, $\varepsilon_i$ is the energy level of the $i$th dot, 
$t_{ii'}$ is the transition probability amplitude between the $i$th and $i'$th dots, 
$U$ is the interdot potential energy, 
$\hbar\Omega_{bk}$ is the energy of the $k$th mode in the $b$th reservoir, 
and $\xi_{ibk}$ is the transition probability amplitude 
between the $i$th dot and the $k$th mode in the $b$th reservoir.
In the second and third terms in the RHS of Eq.~(\ref{HamiltonianDot}), 
the sum is taken over the neighboring dots.
The creation $\hat{d}_i^\dag$ ($\hat{c}_{bk}^\dag$) and annihilation $\hat{d}_i$ ($\hat{c}_{bk}$) operators 
of an electron in the $i$th dot ($k$th mode in the $b$th reservoir) satisfies 
the canonical anti-commutation relations: $\{\hat{d}_i^\dag,\hat{d}_{i'}\}=\delta_{ii'}$, 
$\{\hat{d}_i^\dag,\hat{d}_{i'}^\dag\}=\{\hat{d}_i,\hat{d}_{i'}\}=0$, 
$\{\hat{c}_{bk}^\dag,\hat{c}_{b'k'}\}=\delta_{bb'}\delta_{kk'}$, 
and $\{\hat{c}_{bk}^\dag,\hat{c}_{b'k'}^\dag\}=\{\hat{c}_{bk},\hat{c}_{b'k'}\}=0$.
We assume that the $b$th reservoir is in the equilibrium state 
with inverse temperature $\beta_b$ and chemical potential $\mu_b$ ($b=1,2,...,N_{\rm B}$).
Note that the control parameters are $\{\varepsilon_i\}, \{t_{ii'}\}, U$ (system parameters) and 
$\{\beta_b, \mu_b\}$ (reservoir parameters) in this model.

\subsection{RWA analysis of noninteracting system ($U=0$)}\label{sec:RWANonInt}

We first analyze the noninteracting case ($U=0$) using the RWA.
In this case, by using a linear transformation of the operators $\hat{d}_i^\dag,\hat{d}_i$, 
\begin{align}
\hat{d}_i^\dag &= \sum_{j=1}^N W_{ij}^* \hat{a}_j^\dag, 
\\
\hat{d}_i &= \sum_{j=1}^N W_{ij} \hat{a}_j, 
\end{align}
we can diagonalize the system Hamiltonian $\hat{H}_{\rm S}$:
\begin{align}
\hat{H}_{\rm S} = \sum_{j=1}^N \hbar\omega_j \hat{a}_j^\dag \hat{a}_j,
\end{align}
where $\hbar\omega_j$ is the $j$th mode energy of the noninteracting system.
We can also write the eigenstate $|E_\nu\rangle$ of $\hat{H}_{\rm S}$ as 
$|E_\nu\rangle = \bigotimes_j |\nu_j\rangle$. 
Here, $|\nu_j\rangle$ is either of the empty state $|0_j\rangle$ or singly-occupied state $|1_j\rangle$ 
in the $j$th mode Hilbert space ($\hat{a}_j |0_j\rangle =0$ and  $|1_j\rangle = \hat{a}_j^\dag |0_j\rangle$).

By the above transformation, we can rewrite the system-reservoir coupling Hamiltonian as 
\begin{align}
\hat{H}_{{\rm S}b} = \sum_k \sum_{j=1}^N (\zeta_{jbk} \hat{a}_j^\dag \hat{c}_{bk} + {\rm h.c.}), 
\label{H_Sb_nonIntModel}
\end{align}
where $\zeta_{jbk} = \sum_i W_{ij}^* \xi_{ibk}$.
Note that $\zeta_{jbk}$ depends on the control parameters 
although $\xi_{ibk}$ is not included in the control parameters, 
because $W_{ij}$ depends on $\{\varepsilon_i\}$ and $\{t_{ii'}\}$. 

Now we take the correspondence between the present model and the generic model in Sec.~\ref{sec:setup}.
Equation~(\ref{H_Sb_nonIntModel}) is in the form of Eq.~(\ref{H_Sb_explicit}), 
where $\hat{X}_{b,l} \to \hat{a}_j$ and $\hat{B}_{b,l} \to \sum_k \zeta_{jbk}\hat{c}_{bk}$.
The spectral functions of the reservoir given in Eqs.~(\ref{Phi+}) and (\ref{Phi-}) read 
\begin{align}
\Phi_{b,jj'}^+(\omega) 
= 2\pi \sum_k |\zeta_{jbk}|^2 \delta(\Omega_{bk}-\omega) f^+_b(\omega) \delta_{jj'}, 
\label{Phi+_nonInt}
\\
\Phi_{b,jj'}^-(\omega) 
= 2\pi \sum_k |\zeta_{jbk}|^2 \delta(\Omega_{bk}-\omega) f^-_b(\omega) \delta_{jj'},
\label{Phi-_nonInt}
\end{align}
where 
$f^+_b(\omega) = 1/(1+e^{\beta_b (\hbar\omega - \mu_b)})$ is the Fermi distribution function 
and $f^-_b(\omega) = 1 - f^+_b(\omega)$.
Note that $\Phi_{b,jj'}^\pm(\omega)$ depends on the system parameters as well as the reservoir parameters 
since $\zeta_{jbk}$ depends on $\{\varepsilon_i\}$ and $\{t_{ii'}\}$.

As is mentioned in the previous section, 
it is sufficient to investigate $\mathcal{P}\mathcal{K}_{\rm r}\mathcal{P}$ 
to calculate the excess entropy production $\langle \sigma \rangle^{\rm ex}$ for quasistatic operations.
Furthermore, because $\hat{H}_{\rm S}$ is non-degenerate in the present model, 
each eigenspace is spanned only by a single energy eigenstate $|E_\nu\rangle$.
Using the above-mentioned facts, 
we obtain the matrix representation of the GQMME generator for the noninteracting model within the RWA
in the following form: 
\begin{align}
\Bigl[ \langle E_\kappa | \bigl( \mathcal{K}^\chi_{\rm r} | E_\nu \rangle \langle E_\nu | \bigr) 
| E_\kappa \rangle \Bigr] 
= \sum_{j=1}^N \underbrace{I_2 \otimes \cdots \otimes I_2}_{j-1} \otimes 
\bigl[\mathcal{K}^\chi_{j,\rm r}\bigr] \otimes \underbrace{I_2 \otimes \cdots \otimes I_2}_{N-j}, 
\end{align}
where $I_2$ is the $2 \times 2$ identity matrix, and 
\begin{align}
\bigl[\mathcal{K}^\chi_{j,\rm r}\bigr] = -\frac{v}{\hbar^2}
\begin{pmatrix}
\sum_b \Phi_{b,jj}^+(\omega_j) & -\sum_b \Phi_{b,jj}^-(\omega_j) e^{-i\chi\beta_b(\hbar\omega_j-\mu_b)}
\\
-\sum_b \Phi_{b,jj}^+(\omega_j) e^{i\chi\beta_b(\hbar\omega_j-\mu_b)} & \sum_b \Phi_{b,jj}^-(\omega_j)
\end{pmatrix}.
\label{generatorRWA_matrix_j}
\end{align}
Accordingly we can also decompose $\hat{\ell}_{0,\rm r}^\chi$ and $\hat{r}_{0,\rm r}^\chi$ 
as $\hat{\ell}_{0,\rm r}^\chi = \bigotimes_j \hat{\ell}_{0,j}^\chi$ 
and $\hat{r}_{0,\rm r}^\chi = \bigotimes_j \hat{r}_{0,j}^\chi$, 
where $\hat{\ell}_{0,j}^\chi$ and $\hat{r}_{0,j}^\chi$ are respectively the left and right eigenvectors 
of $\mathcal{K}^\chi_{j,\rm r}$ corresponding to the eigenvalue with the maximum real part. 
By diagonalizing Eq.~(\ref{generatorRWA_matrix_j}), we obtain
\begin{align}
\begin{pmatrix}
\langle 0_j | \hat{\ell}_{0,j}^\chi | 0_j \rangle \\
\langle 1_j | \hat{\ell}_{0,j}^\chi | 1_j \rangle \\
\end{pmatrix}
&= 
\begin{pmatrix}
1 \\ w_j^+(\chi)
\end{pmatrix},
\\
\begin{pmatrix}
\langle 0_j | \hat{r}_{0,j}^\chi | 0_j \rangle \\
\langle 1_j | \hat{r}_{0,j}^\chi | 1_j \rangle \\
\end{pmatrix}
&= C_j(\chi) 
\begin{pmatrix}
1 \\ w_j^-(\chi)
\end{pmatrix},
\label{rightEigenstate_nonInt}
\end{align}
where 
\begin{align}
w_j^\pm(\chi) &= \frac{\sum_b \bigl\{ \Phi_{b,jj}^+(\omega_j)-\Phi_{b,jj}^-(\omega_j) \bigr\} + \sqrt{D_j(\chi)}}
{2 \sum_b \Phi_{b,jj}^\pm(\omega_j) e^{\pm i\chi\beta_b(\hbar\omega_j-\mu_b)}},
\\
D_j(\chi) &= \biggl[ \sum_b \bigl\{ \Phi_{b,jj}^+(\omega_j)-\Phi_{b,jj}^-(\omega_j) \bigr\} \biggr]^2 
+ 4 \sum_{bb'} \Phi_{b,jj}^+(\omega_j) \Phi_{b',jj}^-(\omega_j) 
e^{i\chi[\beta_b(\hbar\omega_j-\mu_b)-\beta_{b'}(\hbar\omega_j-\mu_{b'})]}.
\end{align}
From the normalization condition for $\chi =0$, ${\rm Tr}_{\rm S} \hat{r}_{0,j}^{\chi=0}=1$, 
we have $C_j(0) = \sum_b \Phi_{b,jj}^-(\omega_j) / \gamma_j(\omega_j)$, 
where 
$\gamma_j(\omega) := \sum_b \gamma_{bj}(\omega)$ 
and $\gamma_{bj}(\omega) := \Phi_{b,jj}^+(\omega) + \Phi_{b,jj}^-(\omega)$.
We thus obtain the vector potential for the excess entropy production in Eq.~(\ref{geometricalExcess}) 
for this noninteracting model: 
\begin{align}
\bm{A}_{\rm r}(\bm{\alpha}) 
&= - \sum_{j=1}^N \frac{\partial w_j^+(\chi)}{\partial (i\chi)} \bigg|_{\chi=0} 
\frac{\partial \bigl( C_j(0)w_j^-(0) \bigr)}{\partial \bm{\alpha}} 
\notag\\
&= \sum_{j=1}^N \frac{\sum_b \beta_b(\hbar\omega_j - \mu_b) \gamma_{bj}(\omega_j)}{\gamma_j(\omega_j)}
\frac{\partial}{\partial \bm{\alpha}} \left( \frac{\sum_b \Phi_{b,jj}^+(\omega_j)}{\gamma_j(\omega_j)} \right). 
\label{vectorPotential_nonInt}
\end{align}
We can show $\gamma_{bj}(\omega) = 2\pi \sum_k |\zeta_{jbk}|^2 \delta(\Omega_{bk}-\omega)$ 
and $\Phi_{b,jj}^\pm(\omega) = \gamma_{bj}(\omega) f_b^\pm(\omega)$ 
with the aid of Eqs.~(\ref{Phi+_nonInt}) and (\ref{Phi-_nonInt}).

As a special case of the external operations, we investigate the quasistatic operation 
where we modulate the parameters of only one of the reservoirs, say $\beta_{\rm L}$ and $\mu_{\rm L}$.
In this case, we can write the vector potential (\ref{vectorPotential_nonInt}) 
as the derivative of a scalar function $S_{\rm eff}$ of $\beta_{\rm L}$ and $\mu_{\rm L}$. 
That is, 
\begin{align}
- {\rm Tr}_{\rm S} \left[ \hat{\ell}_{0,\rm r}^{\prime\dag} 
\frac{\partial \hat{\rho}_{\rm ss,r}}{\partial \alpha_{\rm L}} \right] 
&= \frac{\partial S_{\rm eff}(\beta_{\rm L},\mu_{\rm L})}{\partial \alpha_{\rm L}} 
\label{scalarPotentialDescription}
\end{align}
holds for $\alpha_{\rm L} = \beta_{\rm L}, \mu_{\rm L}$, where 
\begin{align}
S_{\rm eff}(\beta_{\rm L},\mu_{\rm L}) 
= - \sum_{j=1}^N \sum_b \sum_{s=\pm} 
\frac{\gamma_{{\rm L}j}(\omega_j) \gamma_{bj}(\omega_j) f_{\rm L}^s (\omega_j)}
{\gamma_j(\omega_j)^2}
\ln f_b^s (\omega_j)
\label{scalarPotential}
\end{align}
Therefore, for this special case of the quasistatic operation in the noninteracting model, 
the excess entropy production is written as the difference of the initial and final values 
of the scalar function $S_{\rm eff}$.
We note that this scalar function $S_{\rm eff}$ is not equal to 
the von Neumann entropy $S_{\rm vN}(\hat{\rho}_{\rm ss,r})$ of the steady state
($S_{\rm vN}(\hat{\rho}_{\rm ss,r}) = S_{\rm sym}(\hat{\rho}_{\rm ss,r})$ for non-degenerate $\hat{H}_{\rm S}$), 
which is given by 
\begin{align}
S_{\rm vN}(\hat{\rho}_{\rm ss,r}) = - \sum_{j=1}^N \sum_{s=\pm} 
\frac{\sum_b \gamma_{bj}(\omega_j) f_b^s(\omega_j)}{\gamma_j(\omega_j)} 
\ln \left( \frac{\sum_b \gamma_{bj}(\omega_j) f_b^s(\omega_j)}{\gamma_j(\omega_j)} \right).
\end{align}
We can derive this from Eq.~(\ref{rightEigenstate_nonInt}) with $\chi = 0$.
We can also show that in the weakly nonequilibrium regime 
the difference between $S_{\rm eff}$ and $S_{\rm vN}(\hat{\rho}_{\rm ss,r})$ is given by 
$S_{\rm eff} - S_{\rm vN}(\hat{\rho}_{\rm ss,r}) = c_0 + c_1 + O(\epsilon^2)$, 
where $c_0$ and $c_1$ are respectively the $O(\epsilon^0)$ and $O(\epsilon)$ constants 
which are independent of $\beta_{\rm L}$ and $\mu_{\rm L}$.
This is consistent with the extended Clausius equality in this regime.

\subsection{Analysis without RWA}\label{sec:NonRWA}

Next, we numerically analyze the model (\ref{HamiltonianDot})-(\ref{coupleHamiltonianDot}) 
without using the RWA. 
We here restrict our interest to the four-dot system ($i=1,2,3,4$) with two reservoirs ($b={\rm L,R}$), 
as illustrated in Fig.~\ref{fig:example}.
In this case, the coupling constant $\xi_{ibk}$ in Eq.~(\ref{coupleHamiltonianDot}) is given by 
$\xi_{i{\rm L}k} = \xi_{1{\rm L}k} \delta_{i1}$ and $\xi_{i{\rm R}k} = \xi_{4{\rm R}k} \delta_{i4}$.
The spectral functions of the reservoirs are defined as 
$\Gamma_{\rm L}(\omega) = 2\pi \sum_k |\xi_{1{\rm L}k}|^2 \delta(\Omega_{{\rm L}k}-\omega)$
and $\Gamma_{\rm R}(\omega) = 2\pi \sum_k |\xi_{4{\rm R}k}|^2 \delta(\Omega_{{\rm R}k}-\omega)$. 
We also assume the wide band limit and the symmetric coupling, i.e., 
$\Gamma_{\rm L}(\omega) = \Gamma_{\rm R}(\omega) = {\rm const.} =: \Gamma$.%
\footnote{
For $U=0$, the spectral function $\Gamma_b(\omega)$ is related with $\gamma_{bj}(\omega)$ in the previous subsection: 
\begin{align*}
\gamma_{{\rm L}j}(\omega) = |W_{1j}|^2 \Gamma_{\rm L}(\omega), 
\\
\gamma_{{\rm R}j}(\omega) = |W_{4j}|^2 \Gamma_{\rm R}(\omega). 
\end{align*}
Therefore, $\gamma_{bj}(\omega)$ depends on the system parameters even in the wide band limit,
since $W_{ij}$ does.
}

We use the geometrical formula (\ref{geometricalExcess}) 
to obtain the excess entropy production $\langle \sigma \rangle^{\rm ex}$ for the quasistatic operations. 
That is, we numerically solve the eigenvalue problem of the GQMME generator $\mathcal{K}^\chi$, 
calculate the vector potential $\bm{A}(\bm{\alpha})$, 
and integrate it along the curve of the operation in the parameter space.

\subsubsection*{Excess Entropy Production for Modulation of Reservoir Parameters of Single Reservoir}

\begin{figure}[bt]
\begin{center}
\includegraphics[width=0.9\linewidth]{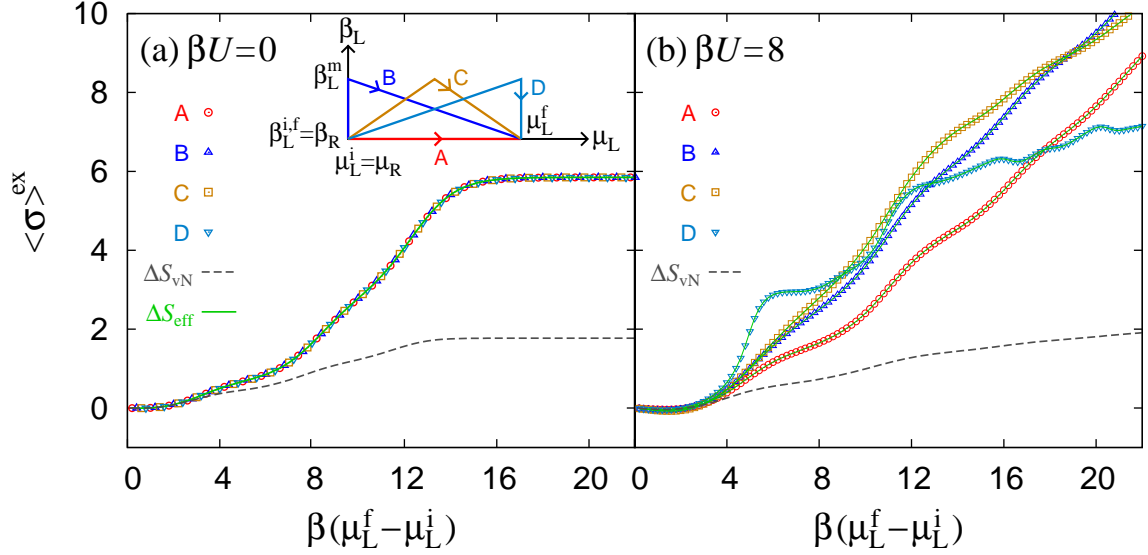}
\caption{
(Color online) Excess entropy production in the four-dot system for a quasistatic operations (A--D) 
of the parameters $\beta_{\rm L}$ and $\mu_{\rm L}$ of the single reservoir L.
Each path starts from the equilibrium condition of $\mu_{\rm L}^{\rm i}=\mu_{\rm R}$ and 
$\beta_{\rm L}^{\rm i} = \beta_{\rm R} = \beta$, goes along the corresponding line 
illustrated in the inset of (a), and ends with the nonequilibrium condition of $\mu_{\rm L}^{\rm f}>\mu_{\rm R}$
and $\beta_{\rm L}^{\rm f} = \beta_{\rm R}$.
The results for a noninteracting ($\beta U = 0$) and an interacting ($\beta U = 8$) systems are plotted 
against $\beta(\mu_{\rm L}^{\rm f}-\mu_{\rm L}^{\rm i})$ in (a) and (b), respectively.
The dashed lines in (a) and (b) are the changes of the von Neumann entropy $S_{\rm vN}$ 
calculated within the RWA.
The solid line in (a) is the change of the function $S_{\rm eff}$ defined in Eq.~(\ref{scalarPotential}).
The solid lines in (b) are $\langle \sigma \rangle^{\rm ex}$ calculated within the RWA.
The parameters are set to $\beta \varepsilon_i =10$ (for all $i$), $\beta t_{ii'} = 4$ (for all $i,i'$), 
$\beta \Gamma= 0.01$, $\beta \mu_{R} = 4$, and $\beta_{\rm L}^{\rm m}/\beta = 5$. 
}
\label{fig:reservoir}
\end{center}
\end{figure}

We here calculate $\langle \sigma \rangle^{\rm ex}$ 
for an operation of the parameters $\beta_{\rm L}$ and $\mu_{\rm L}$ of the reservoir L. 
We set the initial condition to $\mu_{\rm L}^{\rm i}=\mu_{\rm R}$ 
and $\beta_{\rm L}^{\rm i} = \beta_{\rm R} =: \beta$ (equilibrium condition), 
and the final condition to $\mu_{\rm L}^{\rm f}>\mu_{\rm R}$ 
and $\beta_{\rm L}^{\rm f} = \beta_{\rm R}$ (nonequilibrium condition).
We calculate $\langle \sigma \rangle^{\rm ex}$ for four paths (denoted by A, B, C, and D)
connecting these two conditions in the parameter space 
which are illustrated in the inset of Fig.~\ref{fig:reservoir}(a).
For the paths B--D, we set middle conditions to $\beta_{\rm L}^{\rm m}>\beta_{\rm R}$. 
In Fig.~\ref{fig:reservoir}, 
we plot the results as a function of the difference between the initial and final values of $\mu_{\rm L}$ 
(with fixing the value of $\beta_{\rm L}^{\rm m}$). 
We show the results for noninteracting ($\beta U =0$) and interacting ($\beta U =8$) systems 
in Figs.~\ref{fig:reservoir}(a) and (b), respectively.

In Fig.~\ref{fig:reservoir}(a), we observe that the data of $\langle \sigma \rangle^{\rm ex}$ 
for all of the paths agree in the whole range of $\mu_{\rm L}^{\rm f}-\mu_{\rm L}^{\rm i}$. 
Furthermore, the change $\Delta S_{\rm eff}$ of the scalar function given in Eq.~(\ref{scalarPotential}) 
(plotted as a solid line), quantitatively agrees with these data. 
These results indicates that the statement in the RWA analysis on the noninteracting system 
described around Eqs.~(\ref{scalarPotentialDescription}) and (\ref{scalarPotential}) 
in the previous subsection is valid even in the non-RWA analysis.

For interacting systems, this statement is not valid. 
We can clearly see this breakdown in Fig.~\ref{fig:reservoir}(b), 
where the results for the different paths show different behaviors 
in the range of large $\beta(\mu_{\rm L}^{\rm f}-\mu_{\rm L}^{\rm i})$ ($\gtrsim4$).
This result suggests that the situations where $\bm{A}$ is given by the gradient of a scalar potential 
in the strongly nonequilibrium regime are quite exceptional.

In contrast, for small $\beta(\mu_{\rm L}^{\rm f}-\mu_{\rm L}^{\rm i})$ ($\lesssim1$), 
the results for all of the paths are nearly equal. 
Moreover, in this range, these results almost agree with 
the change of the von Neumann entropies $S_{\rm vN}(\hat{\rho}_{\rm ss,r})$ 
between the initial and final steady states (calculated within the RWA and plotted as dashed lines) 
both in Figs.~\ref{fig:reservoir}(a) and (b). 
These observations are consistent with the fact 
that the extended Clausius equality holds in the weakly nonequilibrium regime.

In Fig.~\ref{fig:reservoir}(b), we also show the results for the interacting system analyzed in the RWA 
(plotted as solid lines). 
To obtain these data, we numerically solve the eigenvalue problem 
of the RWA-GQMME generator $\mathcal{K}^\chi_{\rm r}$ in stead of the non-RWA-GQMME generator $\mathcal{K}^\chi$.
We see that all of the data agree with those without the RWA for the corresponding paths.
This result is consistent with the equivalence between the RWA and non-RWA shown in Sec.~\ref{sec:main}.

\subsubsection*{Excess Entropy Production for Cycle Process of Dot Energy Levels}

\begin{figure}[bt]
\begin{center}
\includegraphics[width=0.6\linewidth]{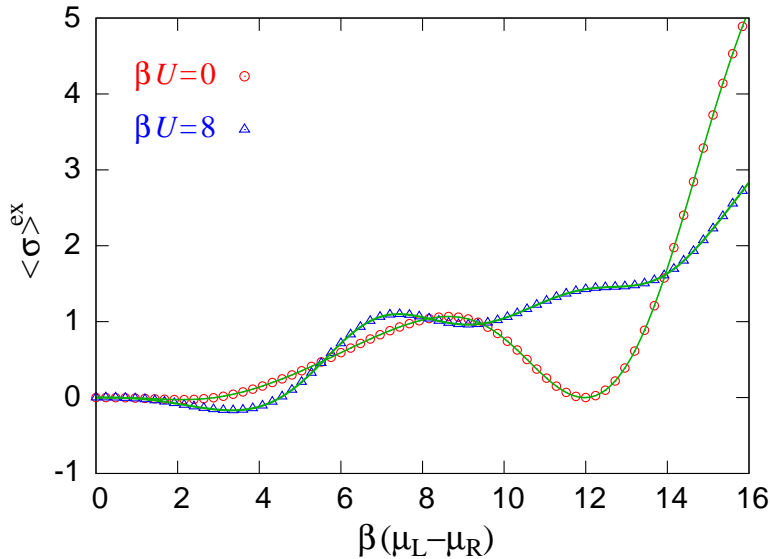}
\caption{
(Color online) Excess entropy production in the four-dot system 
for a quasistatic cycle operation of the dot energy levels 
under the nonequilibrium condition of $\mu_{\rm L} > \mu_{\rm R}$ with $\beta_{\rm L}=\beta_{\rm R}=\beta$ 
[see Eqs.~(\ref{cycleL}) and (\ref{cycleR}) for the detail of the operation].
The results for a noninteracting ($\beta U = 0$; circles) and an interacting ($\beta U = 8$; triangles) 
systems are plotted against $\beta(\mu_{\rm L} - \mu_{\rm R})$. 
The solid lines are the results analyzed within the RWA.
The parameters are set to $\beta \varepsilon_c =10$, $\beta \varepsilon_r =3$, 
$\beta t_{ii'} = 4$ (for all $i,i'$), $\beta \Gamma= 0.01$, and $\beta \mu_{R} = 4$. 
}
\label{fig:cycle}
\end{center}
\end{figure}

We next investigate the excess entropy production for system parameter operations.
In Fig.~\ref{fig:cycle}, we plot $\langle \sigma \rangle^{\rm ex}$ 
for the cycle operations of the dot energy levels
under a nonequilibrium condition of $\Delta\mu := \mu_{\rm L} - \mu_{\rm R}>0$ 
(and the same temperature condition $\beta_{\rm L}=\beta_{\rm R}=:\beta$).
We modulate the energy levels along the circle given by 
\begin{align}
\varepsilon_1 &= \varepsilon_2 = \varepsilon_c + \varepsilon_r \cos\phi, 
\label{cycleL}
\\
\varepsilon_3 &= \varepsilon_4 = \varepsilon_c + \varepsilon_r \sin\phi, 
\label{cycleR}
\end{align}
with $\phi \in [0,2\pi)$.

We observe that $\langle \sigma \rangle^{\rm ex} \simeq 0$ in the weakly nonequilibrium regime 
($\beta\Delta\mu \lesssim 1$) 
for both the noninteracting ($\beta U=0$) and interacting ($\beta U =8$) systems. 
This indicates that the extended Clausius equality is valid in this regime.
In the strongly nonequilibrium regime ($\beta\Delta\mu \gtrsim 2$), in contrast, 
$\langle \sigma \rangle^{\rm ex}$ takes nonzero values, which implies the failure of 
the extension of the Clausius equality with the excess entropy production (even in the noninteracting system). 

We also show the results analyzed in the RWA in Fig.~\ref{fig:cycle} (plotted as solid lines).
To obtain these results, we use Eq.~(\ref{vectorPotential_nonInt}) for the noninteracting system, 
whereas for the interacting system we numerically solve the eigenvalue problem 
of the RWA-GQMME generator $\mathcal{K}^\chi_{\rm r}$ in stead of the non-RWA-GQMME generator $\mathcal{K}^\chi$.
We see that the data within the RWA agree with those without the RWA in the whole range of $\Delta\mu$.
This result is consistent with the equivalence between the RWA and non-RWA shown in Sec.~\ref{sec:main}.

\section{Summary and Discussions}\label{sec:summary}

For open quantum systems described by the QMME, 
we have derived a geometrical expression for the excess entropy production 
during an arbitrary quasistatic operation that connects two NESSs.
In the derivation, we have used the technique of the full counting statistics and the adiabatic approximation.
This result implies that the scalar thermodynamic potential for arbitrary NESSs cannot be defined 
from the excess entropy production for the quasistatic operation, 
and that the vector potential $\bm{A}$ 
plays a crucial role in the steady state thermodynamics (SST).

We have also shown that the result of the excess entropy production within the RWA 
is equivalent to that without the RWA. 
This is helpful for the investigation of the SST in the framework of the QMME, 
because the RWA makes calculation easier 
(in particular, if the system Hamiltonian is non-degenerate, 
the form of the QMME is equivalent to that of the classical Markov jump process).

In the weakly nonequilibrium regime, with the aid of the RWA, 
we have derived the extended Clausius equality from the geometrical expression.
This result extends the validity range of the equality 
derived in classical systems \cite{KNST1,KNST2,SagawaHayakawa} 
and quantum heat conducting systems \cite{SaitoTasaki}
to the systems described by the QMME (including electrical conducting systems).

As an example, we have investigated a spinless electron system in quantum dots. 
We have found that in the noninteracting systems 
there exists a scalar potential for the operation on a single reservoir, 
but that this is not valid in the interacting systems.

There are many issues to be studied in the future. 
One of the important issues is the way of constructing a thermodynamic potential in the SST.
We are considering two directions. 
One is that we construct the thermodynamics by using the vector potential as the thermodynamic potential. 
For this purpose, it is important to investigate the thermodynamic structure from the geometrical viewpoint 
\cite{Zulkowski}.
The other is that we restrict the class of systems of the SST to ``macroscopic'' systems. 
By the restriction, it may be possible to construct a scalar potential from the excess entropy production.

\bigskip
{\small 
The authors thank Keiji Saito for his helpful advice. 
This work was supported by 
a JSPS Research Fellowship for Young Scientists (No. 24-1112),  
a Grant-in-Aid for Research Activity Start-up (KAKENHI 11025807), 
and a Grant-in-Aid (KAKENHI 25287098). 
A part of this study was performed when TY and TS were in the Yukawa Institute for Theoretical Physics.
}


\end{document}